\documentclass[runningheads]{llncs}

\usepackage{eccv}

\usepackage{eccvabbrv}

\usepackage{graphicx}
\usepackage{booktabs}

\usepackage[accsupp]{axessibility}  

\usepackage{booktabs}
\usepackage{multirow}
\usepackage{amsmath}
\usepackage{amsfonts}
\usepackage{xcolor}
\usepackage{colortbl}
\usepackage{amssymb}
\usepackage{pifont}

\definecolor{mygreen}{RGB}{52,121,40}
\definecolor{lightblue}{RGB}{230, 244, 252}

\newcommand{\cmark}{\ding{51}}%
\newcommand{\xmark}{\ding{55}}%

\newcommand{\tinygreenup}[1]{{\tiny\color{mygreen}($\uparrow$#1\%)}}

\usepackage{hyperref}

\usepackage{orcidlink}

\begin{document}

\title{Precise Video-to-Audio Generation with Cross-Modal Alignment in Latent Space} 

\titlerunning{Flowley}

\author{Thanh V. T. Tran\inst{1} \and
Ngoc-Son Nguyen\inst{1} \and
Luong Tran\inst{1} \and
Long-Khanh Pham\inst{1} \and
Paarth Neekhara\inst{2} \and
Shehzeen Hussain\inst{2} \and
Van Nguyen\inst{1}} 

\authorrunning{Tran et al.}

\institute{FPT Software AI Center, Vietnam \and
NVIDIA Corporation, USA \\
\url{https://flowley-v2a.github.io} \\
\email{\{thanhtvt1,sonnn45,luongtk,khanhpl2,vannth19\}@fpt.com}
\email{\{pneekhara,shehzeenh\}@nvidia.com}}

\maketitle

\begin{abstract}
Video-to-audio (V2A) generation aims to synthesize realistic audio that is both semantically consistent with and temporally synchronized to a silent video. Despite recent progress, many methods still rely on multi-stage training, resulting in high computational costs and long runtimes, or transform visual input into text to leverage pretrained text-to-audio models, sacrificing fine-grained temporal cues. To overcome these limitations, we propose Flowley, an end-to-end, single-stage training architecture that produces soundtracks by combining visual features with textual prompts. Crucially, we introduce Progressive Soft-masked Cross-Attention, which embeds audio-visual synchronization directly within its attention mechanism, adding zero additional computational cost compared to standard attention layers. We further observe that existing V2A benchmarks lack sound-oriented descriptive captions, which can potentially degrade the quality of the synthesized audio. To remedy this, we propose SoundCap, a plug-and-play pipeline for creating detailed, sound-aware captions that guide the model. Remarkably, without integrating any pretrained audio-visual alignment modules, Flowley achieves state-of-the-art performance on VGGSound across multiple metrics. Moreover, by incorporating SoundCap, we further exceed the performance of the strongest existing close-sourced methods in terms of audio quality in the zero-shot setting. 
\end{abstract}

\section{Introduction}
\label{sec:intro}

Sound design is the craft of storytelling through sonic composition. A key branch of this field is Foley \cite{ThemeAment2014}, where sound effects are created in precise synchronization with onscreen action during post-production. These soundscapes are brought to life by a skilled Foley artist working on a purpose-built stage stocked with a wide array of props and materials for generating the required sounds\footnote{See how a foley artist performing a scene in our project page.}.

Recent advances in generative audio modeling have facilitated text-to-audio (T2A) systems \cite{10.1109/TASLP.2024.3399607,10.1145/3664647.3681688,10888461}, allowing designers to generate sound effects directly from written descriptions. Despite accelerating the search for appropriate sounds, they must still manually adjust timing to align audio with on-screen action. This contrasts sharply with traditional Foley work, where artists naturally craft and sync sound effects in real time by interacting physically with props.

To overcome this limitation, video-to-audio (V2A) generation has gained significant attention, dividing into two major directions. The first converts visual features from silent video frames into text, leveraging pretrained T2A models to synthesize sound \cite{Wang_Ma_Pascual_Cartwright_Cai_2024,zhang2024foleycrafterbringsilentvideos}. While this approach leverages the strengths of pretrained T2A systems, it inevitably discards fine-grained temporal details, which are crucial for film-production sound effects. The second line of work uses multi-stage training: early phases learn to extract or align acoustic cues from video frames, either via dedicated regression networks \cite{Jeong_Kim_Chun_Lee_2025, Wang_2025_CVPR, ton2025taro} or contrastive objectives \cite{NEURIPS2023_98c50f47}, and subsequent stages build on these pretrained modules. While effective, these pipelines incur substantial delays and computational overhead. Furthermore, integrating textual descriptions into the generation workflow, beyond their use in T2A systems, remains under-explored, despite the demonstrated benefits of text guidance in other domains \cite{Kirillov_2023_ICCV,Wang_Qi_Wang_Sun_Zhuang_Wu_Zhang_Liao_2025,10888112}. This gap is partly due to existing V2A datasets \cite{Owens_2016_CVPR,9053174} lacking rich, sound-focused text annotations.

In this work, we introduce \textit{Flowley}, a flow-based architecture that synthesizes high-fidelity audio from silent video by harnessing multimodal cues. Our framework consists of two interconnected modules: \textbf{(1)} multi-stream blocks, which jointly fuse audio-latent, visual, and textual embeddings into rich, cross-modal representations; and \textbf{(2)} single-stream blocks, which further refine the audio pathway. To enhance multimodal conditioning within each single-stream block, we incorporate a cross-attention mechanism with textual features and propose a novel \textit{\textbf{P}rogressive \textbf{S}oft-masked \textbf{C}ross-\textbf{A}ttention} (PSCA) module that effectively aligns acoustic and visual information. Empirically, we show that this design improves the temporal alignment of synthesized audio, eliminating the need for pretrained synchronization modules or additional pretraining stages. In practice, however, most V2A datasets either lack descriptive annotations or contain low-quality captions, which limits a model’s ability to maintain semantic consistency in generated audio. To overcome this, we introduce \textit{\textbf{Sound}-aware \textbf{Cap}tioner} (SoundCap), a plug-and-play pipeline that leverages pretrained audio-visual large language models (AV-LLMs) to generate detailed, sound-oriented captions as ground truth for training vision-language models (VLMs), thereby enabling robust inference of both on- and off-screen acoustic events. Comprehensive experiments on VGGSound show that Flowley exceeds state-of-the-art (SOTA) methods across a range of metrics. Notably, when augmented with SoundCap, Flowley surpasses Movie Gen Audio in zero-shot audio quality, underscoring the added effectiveness of SoundCap.

\section{Related Works}

\subsection{Video-to-Audio Generation}

V2A generation has garnered significant interest because of its potential utility in multimedia production. Early efforts predominantly employed autoregressive models: Zhou \etal \cite{Zhou_2018_CVPR} first demonstrated SampleRNN \cite{mehri2017samplernn} for direct waveform synthesis from video frames, while SpecVQGAN \cite{SpecVQGAN_Iashin_2021} used a transformer-based autoregressive architecture conditioned on RGB inputs and optical flow. Im2Wav \cite{10096023} further advanced this line by adopting a dual-resolution transformer framework and leveraging CLIP \cite{pmlr-v139-radford21a} embeddings to predict VQVAE code indices. However, these autoregressive methods are hampered by slow inference speed, as they must generate audio samples sequentially. To overcome these limitations, diffusion-based approaches have emerged, demonstrating high-quality soundtrack synthesis \cite{pham2025mdsgen}. Luo \etal \cite{NEURIPS2023_98c50f47}, Wang \etal \cite{NEURIPS2024_e7384de3}, and Zhang \etal \cite{zhang2025modelguided} designed an audio-visual contrastive feature and utilized latent diffusion and flow-based model to predict mel-spectrograms. V2A-Mapper \cite{Wang_Ma_Pascual_Cartwright_Cai_2024} projects CLIP visual embeddings into CLAP space for conditioning AudioLDM \cite{pmlr-v202-liu23f}. Temporal conditioning has also been explored in SonicVisionLM \cite{Xie_2024_CVPR}, ReWaS \cite{Jeong_Kim_Chun_Lee_2025}, FoleyCrafter \cite{zhang2024foleycrafterbringsilentvideos}, STA-V2A \cite{10890132}, and Mel-QCD \cite{Wang_2025_CVPR}. However, these frameworks typically presume perfect alignment between text and video and depend on frozen T2A models, which can cause textual cues to dominate visual information. Recently, VinTAGe \cite{Kushwaha_2025_CVPR} and MMAudio \cite{Cheng_2025_CVPR} have investigated joint multimodal conditioning but heavily rely on external pretrained audio-visual alignment modules to ensure synchronization and semantic consistency. In contrast, Flowley achieves SOTA results with a streamlined, single-stage training pipeline that forgoes any external pretrained audio-visual alignment modules.

\subsection{Visual-guided Audio Generation with Acoustic-enhanced Captions}

Although simple captions can effectively guide audio synthesis, many models struggle with complex, detail-rich prompts, a challenge commonly referred to as ``prompt following''~\cite{betker2023improving}. While prompt engineering has been explored in vision generation \cite{betker2023improving,videoworldsimulators2024}, it remains under-investigated in audio generation, and major audio–visual datasets like AudioSet \cite{7952261} and VGGSound offer at best minimal textual descriptions. Few works have attempted to bridge this gap: VATT \cite{NEURIPS2024_b782a346} fine-tunes VLMs to predict audio events from video alone using LTU-generated captions as supervision \cite{gong2024listen}, and another approach employs a multi-model pipeline to extract visual and acoustic cues separately before feeding them into an LLM for caption synthesis \cite{10.1145/3664647.3681472,yuan2024soundvecaps}. However, both strategies share the same flaw of cross-modal inconsistency. For instance, when we replicated the setup of Sound-VECaps \cite{yuan2024soundvecaps}, the acoustic model frequently confused the rumble of a car engine with a lion’s roar. To overcome these limitations, we leverage AV-LLMs to produce detailed, sound-focused captions as unified ground truth, thereby eliminating modality mismatch and effectively capturing both on- and off-screen audio events.

\section{Methodology}

\subsection{Preliminaries}

\subsubsection{Flow Matching.}
We use the flow matching (FM) \cite{lipman2023flow,liu2023flow} framework to train our model. FM enables sampling from the target data distribution by progressively transforming a sample drawn from a prior distribution, such as a standard Gaussian distribution. During training, given an audio latent $x_1$, we randomly sample a timestep $t\in[0,1]$ and a noise vector $x_0\sim\mathcal N(0,1)$, which are then combined to form a training point $x_t$. The objective is to train the model to estimate the velocity vector $v_t=\frac{\mathrm{d}}{\mathrm{d}t}x_t$ which teaches it to ``move'' the sample $x_t$, guiding the transformation of $x_t$ toward the target $x_1$. While there are numerous ways to construct $x_t$, we use a simple linear interpolation:
\begin{equation*}
    x_t = tx_1 + (1-t)x_0.
\end{equation*}
This leads to a constant ground-truth velocity:
\begin{equation*}
        v_t = \frac{\mathrm{d}}{\mathrm{d}t}x_t = x_1 - x_0.
\end{equation*}
Letting $\theta$ denote the model parameters and $c$ the multimodal conditioning inputs, the predicted velocity is expressed as $v_\theta(x_t,t,c)$. We train the model by minimizing the MSE between the predicted and ground-truth velocities:
\begin{equation}\label{eq:fm}
    \mathcal L_{\text{FM}}(\theta) = \mathbb E_{t,x_0,x_1,c}||v_\theta(x_t,t,c)-v_t||^2.
\end{equation}

At inference, we begin by sampling $x_0\sim\mathcal N(0,1)$, and then solve an ordinary differential equation (ODE) using the model's estimated velocity field to obtain the final output $x_1$.

\subsubsection{Audio encoding.} 

Following previous works \cite{pmlr-v202-liu23f,NEURIPS2024_e7384de3}, for computational efficiency, we downsample audio to 16 kHz and convert it to a mel-spectrogram, which is then encoded into a latent variable $x_1 \in \mathbb R^{L_{\text{aud}} \times D_{\text{aud}}}$ by a pretrained variational autoencoder (VAE) \cite{Kingma2014}, where $L_{\text{aud}}$ and $D_{\text{aud}}$ represent the sequence length and VAE bottleneck dimension, respectively. During inference, the predicted latent is decoded back into a mel-spectrogram, and the waveform is finally reconstructed using BigVGAN \cite{lee2023bigvgan}.

\begin{figure}[t]
    \centering
    \includegraphics[width=\linewidth]{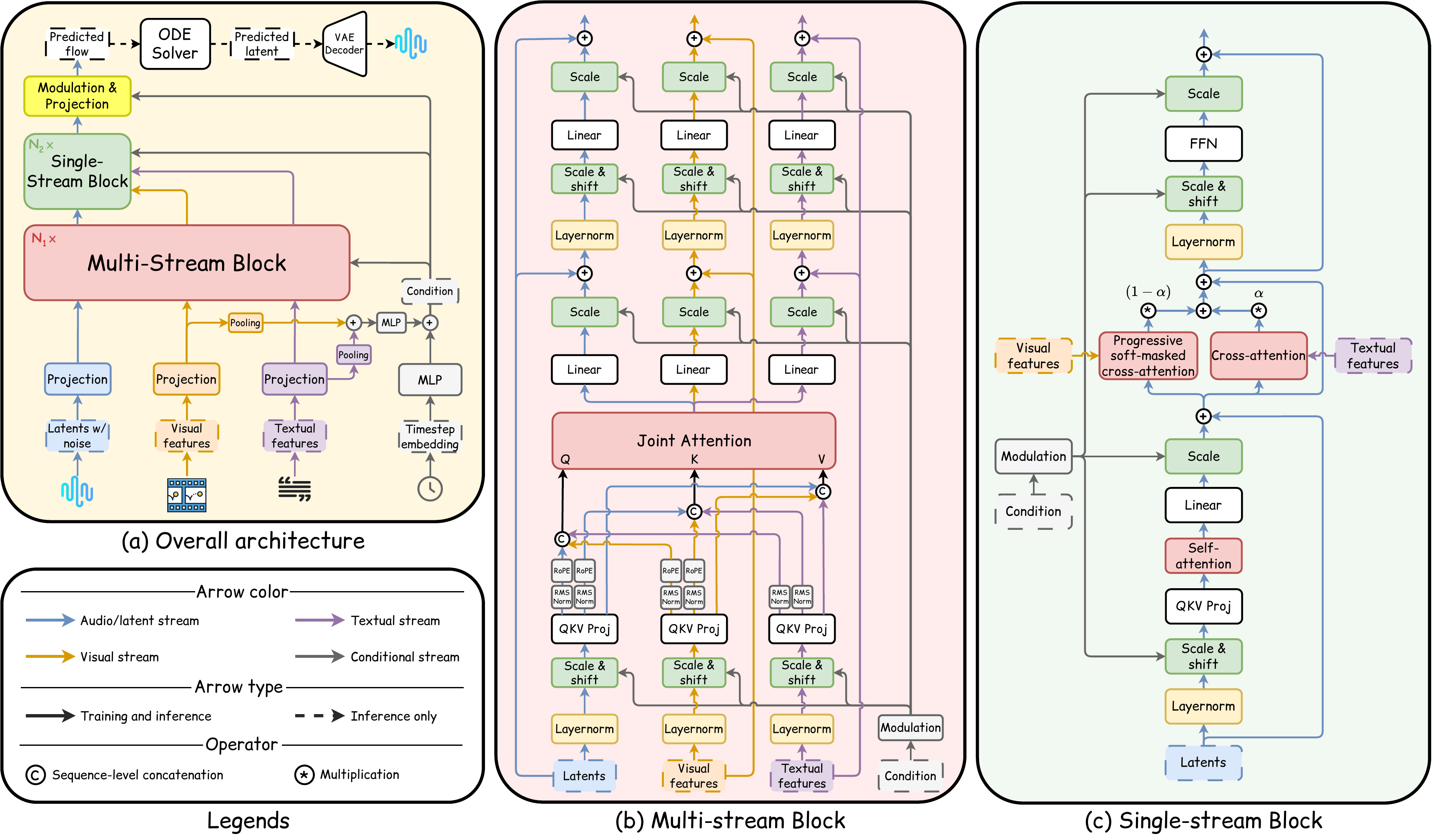}
    \caption{\textbf{(a)} The proposed Flowley framework consists of two core modules. \textbf{(b)} First, visual, textual, and audio latent representations are processed together through the multi-stream block. \textbf{(c)} Latent features are then passed into the single-stream block, where they undergo weighted cross-attention with the visual and textual streams to estimate the flow field. At inference time, we integrate this learned flow using standard ODE solvers to generate the compressed mel-spectrogram, which is subsequently decoded and vocoded into the final audio waveform.}
    \label{fig:flowley}
\end{figure}

\subsection{Flowley's Architecture Overview}

To predict the flow field ${v_\theta}$ for a given latent state ${x}_t$, Flowley incorporates both visual and textual conditioning. As depicted in \Cref{fig:flowley}a, our overall architecture adopts a multi-to-single stream design paradigm \cite{Black_Forest_Labs_2024,Cheng_2025_CVPR}. After obtaining representations from pretrained encoders, Flowley first employs $N_1$ multi-stream blocks \cite{pmlr-v235-esser24a}, which jointly process video, text, and audio latents, and then uses $N_2$ single-stream blocks to refine the audio latent stream, which serves as the primary stream of the block. Crucially, Flowley introduces an audio-visual alignment-injected mechanism that removes the need for any external dedicated modules, which is an aspect that existing approaches have not yet resolved. We describe each of these components below.

\subsubsection{Representations.}

We adopt FLAN-T5 \cite{JMLR:v25:23-0870} as our text encoder to produce embeddings $f_{\text{txt}} \in \mathbb R^{L_{\text{txt}} \times D_{\text{txt}}}$, where $L_{\text{txt}}$ is the number of tokens and $D_{\text{txt}}$ the dimensionality of the textual features. For visual cues, we sample the video at 8 FPS and pass each frame through CLIP \cite{pmlr-v139-radford21a,fang2024data} to obtain frame-wise representations. The resulting video embedding is denoted $f_{\text{vis}} \in \mathbb R^{L_{\text{vis}} \times D_{\text{vis}}}$, where $L_{\text{vis}}$ and $D_{\text{vis}}$ represent the number of frames and visual feature dimension, respectively. Notably, unlike prior methods that rely on pretrained temporal modules, such as onset detectors \cite{zhang2024foleycrafterbringsilentvideos}, energy extractors \cite{Jeong_Kim_Chun_Lee_2025}, or audio-visual alignment networks \cite{NEURIPS2024_e7384de3,Cheng_2025_CVPR}, we integrate temporal alignment directly within our model, as described in the \Cref{sec:psca}. Finally, we combine visual and textual representations with a learned timestep embedding via average pooling and addition to form the global condition $c \in \mathbb R^{1\times D}$, where $D$ is the hidden dimension of Flowley. This condition is injected into the network via adaptive layer normalization (adaLN) layers \cite{Perez_Strub_de_Vries_Dumoulin_Courville_2018}, which apply learned scales and biases. Specifically, given an input $y \in \mathbb R^{L \times h}$, each adaLN layer computes:
\begin{equation*}
    \mathrm{adaLN}(y, c) = \mathrm{LayerNorm}(y)\cdot \mathbf{1}\mathbf{W}_1(c) + \mathbf{1}\mathbf{W}_2(c),
\end{equation*}
where $\mathbf W_1$ and $\mathbf W_2$ are two MLPs, and $\mathbf{1} \in \mathbb R^{L\times 1}$ is an all-ones vector that broadcasts the scale and bias across the sequence length $L$.

\subsubsection{Multi-stream Block.}

Our multi-stream block builds on the MM-DiT image generation architecture \cite{pmlr-v235-esser24a}. As shown in \Cref{fig:flowley}b, it enables cross-modal interaction by applying a joint attention mechanism over visual, textual, and audio latents. To ensure stable training, we follow established best practices \cite{NEURIPS2024_ed2dad59} by integrating QK-Norm and Rotary Positional Embeddings (RoPE) \cite{Su2024RoFormer} directly into the key-query dot-product attention computation.

\subsubsection{Single-stream Block.} After passing through $N_1$ multi-stream blocks, the audio latent is routed into a single-stream network comprising of $N_2$ blocks. This two-stage design helps decouple multimodal fusion from modality-specific refinement, improving scalability and parameter efficiency. For these blocks, we extend the DiT architecture \cite{Peebles_2023_ICCV} by augmenting its cross-attention mechanism to incorporate both learned visual and textual embeddings (see \Cref{fig:flowley}c). By jointly attending to aligned video and audio frames, we enhance temporal synchronization, while the inclusion of detailed, sound-oriented text improves semantic fidelity within the acoustic latent representation. Notably, we introduce PSCA, a novel mechanism that directly aligns visual and acoustic features on a frame-wise basis within the flow model’s latent space. A detailed explanation of PSCA is presented in the next section.

\subsection{Progressive Soft-masked Cross-Attention}\label{sec:psca}

The original cross-attention mechanism, introduced by Vaswani \etal \cite{NIPS2017_3f5ee243}, enables the decoder to focus on the most relevant portions of the encoder’s outputs:
\begin{equation*}
    \mathrm{CA}(Q,K,V) = \mathrm{softmax}\bigg( \frac{QK^T}{\sqrt{d_k}} \bigg)V,
\end{equation*}
where $d_k$ is the hidden dimension of each attention head. In our setting, this translates to acoustic features attending to various frames of the visual input. However, this setup introduces a risk: an audio segment may mistakenly attend to unrelated visual frames, leading to temporal misalignment in the generated audio. To mitigate this, we introduce a soft mask $M$ with entries in $[0,1]$, where values approaching zero impose stronger masking. Specifically, denote the audio query sequence $Q_{\text{aud}} \in \mathbb R^{L_{\text{aud}}\times D}$ and the video key/value sequences $K_{\text{vis}}, V_{\text{vis}} \in \mathbb R^{L_{\text{vis}}\times D}$, sampled at rates $r_a$ and $r_v$ (with $r_a > r_v$). Because each video frame spans multiple audio frames, we define the video frame index aligned to audio position $i$ as:

\begin{equation*}
    j_c(i) = \min\bigg(\bigg\lfloor \frac{r_v}{r_a}i \bigg\rfloor,L_{\text{vis}}-1\bigg).
\end{equation*}
Let $d_{ij} = |j-j_c(i)|$ be the absolute distance (in video-frame indices) between audio token $i$ and video frame $j$, we define our base soft-mask $M_{ij}^{(0)}$ as:
\begin{equation*}
    M_{ij}^{(0)} = \begin{cases}
        1,&d_{ij} \leq \omega,\\
        \mathcal{F}(d_{ij}-\omega),&\omega<d_{ij}\leq \omega+\delta, \\
        0,&d_{ij}>\omega+\delta, \\
    \end{cases}
\end{equation*}
where $\omega$ defines a hard attention window of full weight, and $\delta$ indicates the additional number of video frames beyond the hard window over which attention weights are softly decayed from 1 down to 0, since adjacent frames often carry useful contextual or motion cues. Therefore, instead of discarding them entirely, we apply a decay kernel $\mathcal F: [0, \delta] \rightarrow [0,1]$ that satisfies $\mathcal F(0)=1$ and $\mathcal F(\delta)=0$. The mathematical expression of $\mathcal F$ is as follows:
\begin{equation*}
    \mathcal F(m) = \frac{1}{2}\bigg[ \cos\bigg( \frac{\pi m}{\delta} \bigg) + 1 \bigg].
\end{equation*}

Furthermore, to encourage early layers to explore a wider context while later layers focus on precise local evidence, we scale the fade zone by a layer-dependent progressive parameter $\beta$:
\begin{equation*}
    \beta_\ell = 1 - \frac{\ell}{N_2 - 1},\quad\ell \in \{0,\ldots, N_2-1 \}.
\end{equation*}

Therefore, the final soft-mask for layer $\ell$ is:
\begin{equation*}
    M_{ij}^{(\ell)} = \begin{cases}
        1,&d_{ij} \leq \omega,\\
        \beta_\ell\mathcal{F}(d_{ij}-\omega),&\omega<d_{ij}\leq w+\delta, \\
        0,&d_{ij}>\omega+\delta. \\
    \end{cases}
\end{equation*}

Figure \ref{fig:soft-mask-visualize} visualizes the mask values $M_{ij}^{(\ell)}$ as a function of the audio-video offset $d_{ij}$ and network layer $\ell$. Within any given block, entries closest to the aligned center frame $j_c(i)$ (i.e., smallest $d_{ij}$) exhibit the highest mask values, facilitating stronger cross-modal attention, while more distant frames are progressively attenuated. Moreover, as the layer index increases, the overall mask values diminish, converging to zero in the final layer and effectively reducing PSCA to a hard sliding-window. Given the mask formulation, we formally express the PSCA mechanism at layer $\ell$ as:
\begin{equation*}
    \mathrm{PSCA}_\ell(Q_{\text{aud}},K_{\text{vis}},V_{\text{vis}}) = \mathrm{softmax}\bigg(\frac{Q_{\text{aud}}K_{\text{vis}}^T}{\sqrt{d_k}} + \log(M^{(\ell)} + \epsilon) \bigg)V_{\text{vis}},
\end{equation*}
where $\epsilon$ is a very small number to avoid $\log(0)$. PSCA unifies \textit{hard} locality, \textit{soft} contextual fading, and a \textit{depth-aware} progression in a single, differentiable mask. Early layers benefit from broader, but still weighted, video context, reducing alignment errors, while deeper layers concentrate capacity on the most tightly aligned frames, improving fine-grained synchronization. Finally, the updated representation $\tilde x^{(\ell)}$ of the audio latent stream $x^{(\ell)}$ in $\ell$-th block combines text‐ and vision‐conditioned attention via:
\begin{equation*}
    \begin{split}
        \tilde x^{(\ell)} = x^{(\ell)} + \alpha^{(\ell)} \cdot \mathrm{CA}(Q_{\text{aud}}, K_{\text{txt}}, V_{\text{txt}}) + (1-\alpha^{(\ell)}) \cdot \mathrm{PSCA}_\ell(Q_{\text{aud}}, K_{\text{vis}}, V_{\text{vis}}),
    \end{split}
\end{equation*}
where $\alpha^{(\ell)} \in [0, 1]$ is a learnable parameter determining the contributions from the two cross-attention branches at $\ell$-th block.

\subsection{Training and Inference}

\begin{figure}
\centering
\begin{minipage}[t]{0.5\linewidth}
  \centering
  \includegraphics[width=\textwidth]{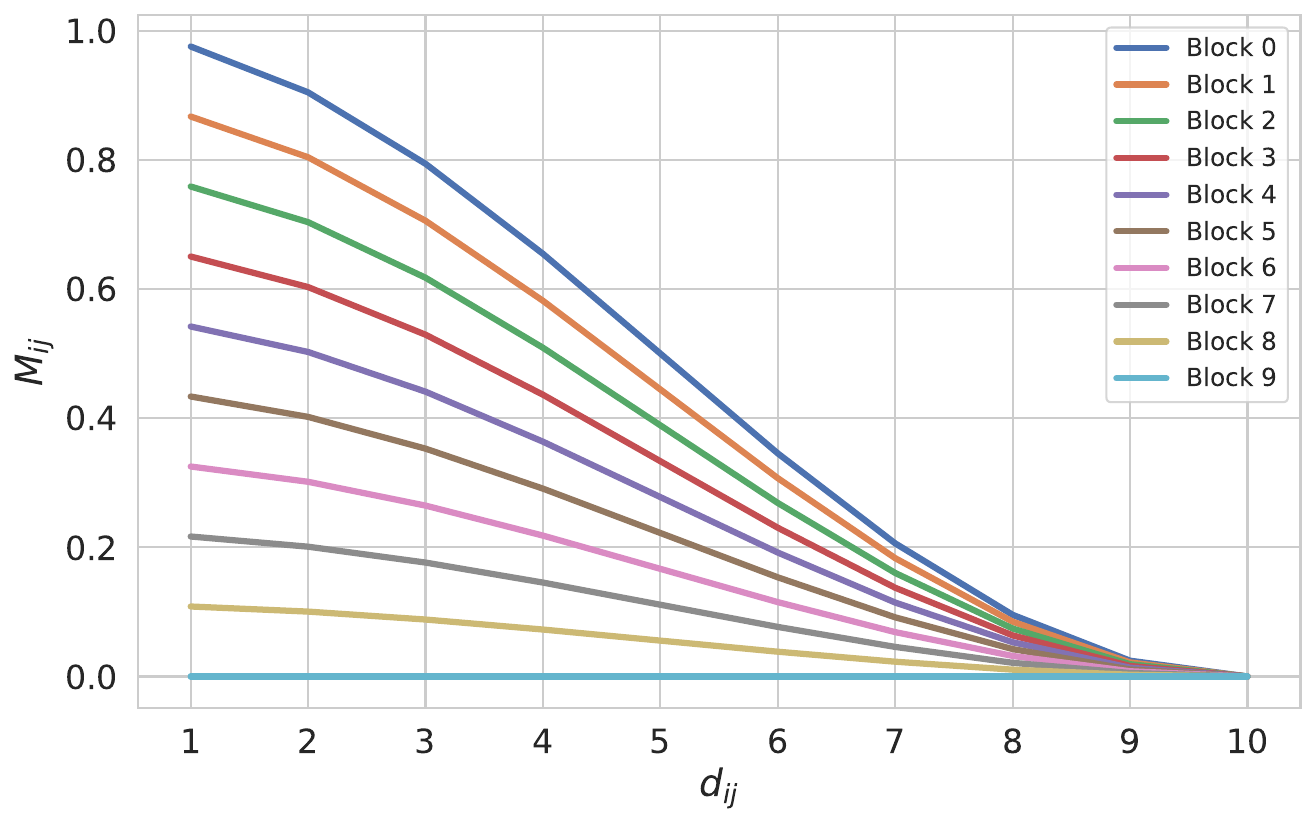}
  \caption{Mask values within the fade region of the progressive soft-mask $M$ across all $N_2=10$ single-stream blocks, using $\omega=0$ and $\delta=10$.}
  \label{fig:soft-mask-visualize}
\end{minipage}%
\hfill
\begin{minipage}[t]{0.48\linewidth}
  \centering
  \includegraphics[width=\textwidth]{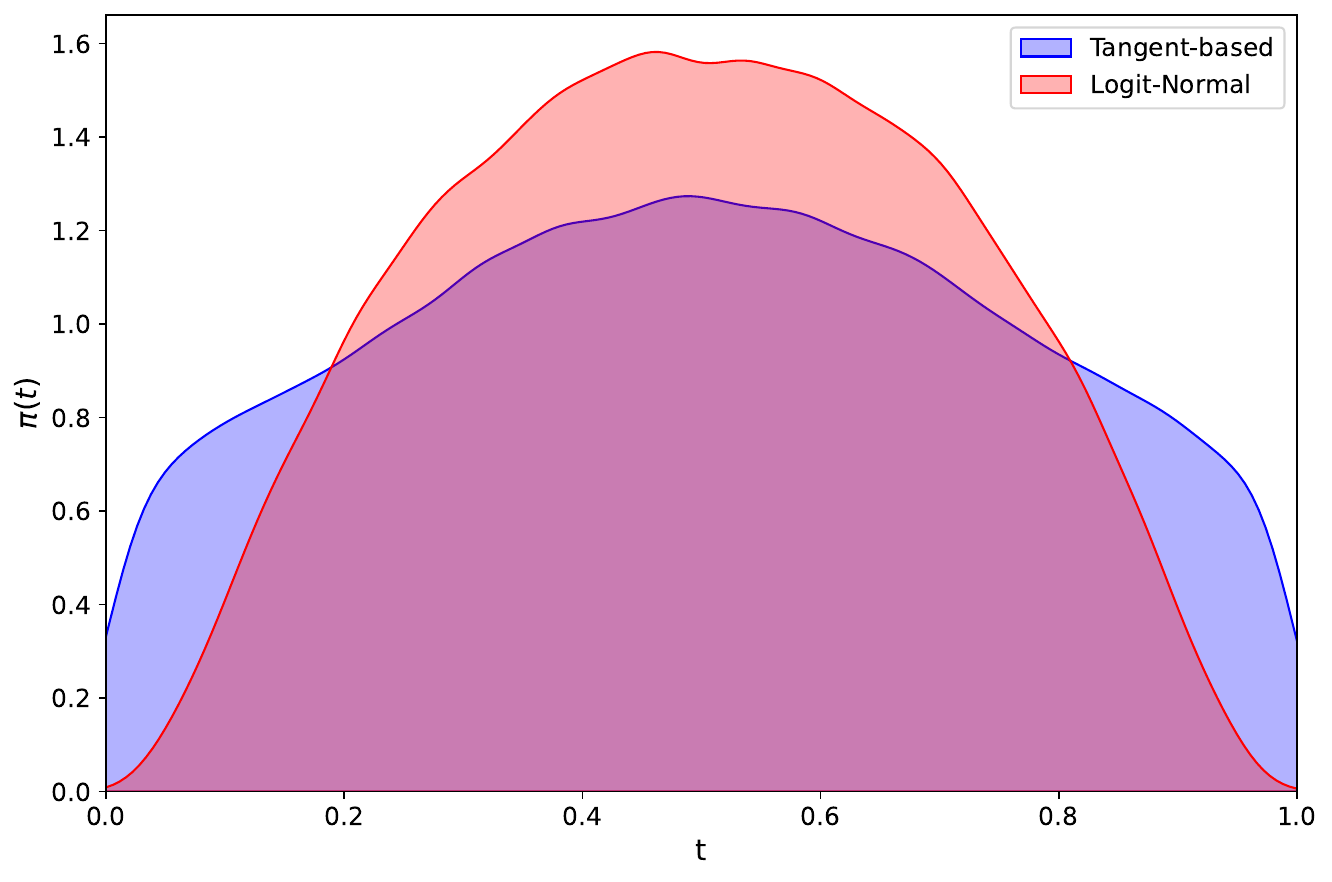}
  \caption{Distribution of tangent-based and logit-normal schedulers.}
  \label{fig:distribution}
\end{minipage}
\end{figure}

The noise scheduler plays a critical role in training both flow-based and diffusion-based models. Although the logit-normal distribution \cite{10.1093/biomet/67.2.261} is often preferred for its emphasis on intermediate timesteps, it collapses to zero density at the endpoints $t=0$ and $t=1$ (\Cref{fig:distribution}). To address this, we employ a tangent-based sampling schedule defined by:
\begin{equation*}
    t = 1-\frac{1}{\tan\Big(\frac{\pi}{2}u\Big) + 1},
\end{equation*}
where $u\sim\mathcal{U}[0,1]$. The resulting distribution is illustrated in~\Cref{fig:distribution} and the mathematical derivation is provided in the Supplementary Material. To accelerate the training process, we leverage data-noise alignment \cite{NEURIPS2024_a422a2f0} which determines in advance the target noise level for each example’s latent before noise injection. Furthermore, in addition to the FM loss (\Cref{eq:fm}), we introduce a velocity direction supervision term using cosine distance:
\begin{equation*}
        \mathcal L_{\text{vel}} = 1 - \text{sim}\big(v_\theta(x_t,t,c), (x_1-x_0)\big),
\end{equation*}
where $\text{sim}(\cdot,\cdot)$ denotes the cosine similarity. Our overall training objective thus becomes $\mathcal L = \mathcal L_{\text{FM}} + \lambda\mathcal{L_{\text{vel}}}$, where $\lambda$ is the hyperparameter. To support classifier-free guidance (CFG) \cite{ho2022classifier} at inference, we randomly mask either the visual tokens or the textual prompt with a $10\%$ probability during training. With that, our velocity prediction becomes:
\begin{equation*}
    \tilde v_\theta(x_t,t,c) = v_\theta(x_t,t,\varnothing) + s\Big(v_\theta(x_t,t,c) - v_\theta(x_t,t,\varnothing)\Big),
\end{equation*}
where $\varnothing$ denotes empty token and $s$ is the guidance weight.

\subsection{Sound-aware Captioner}\label{sec:soundcap}

\begin{figure}[t]
    \centering
    \includegraphics[width=0.8\linewidth]{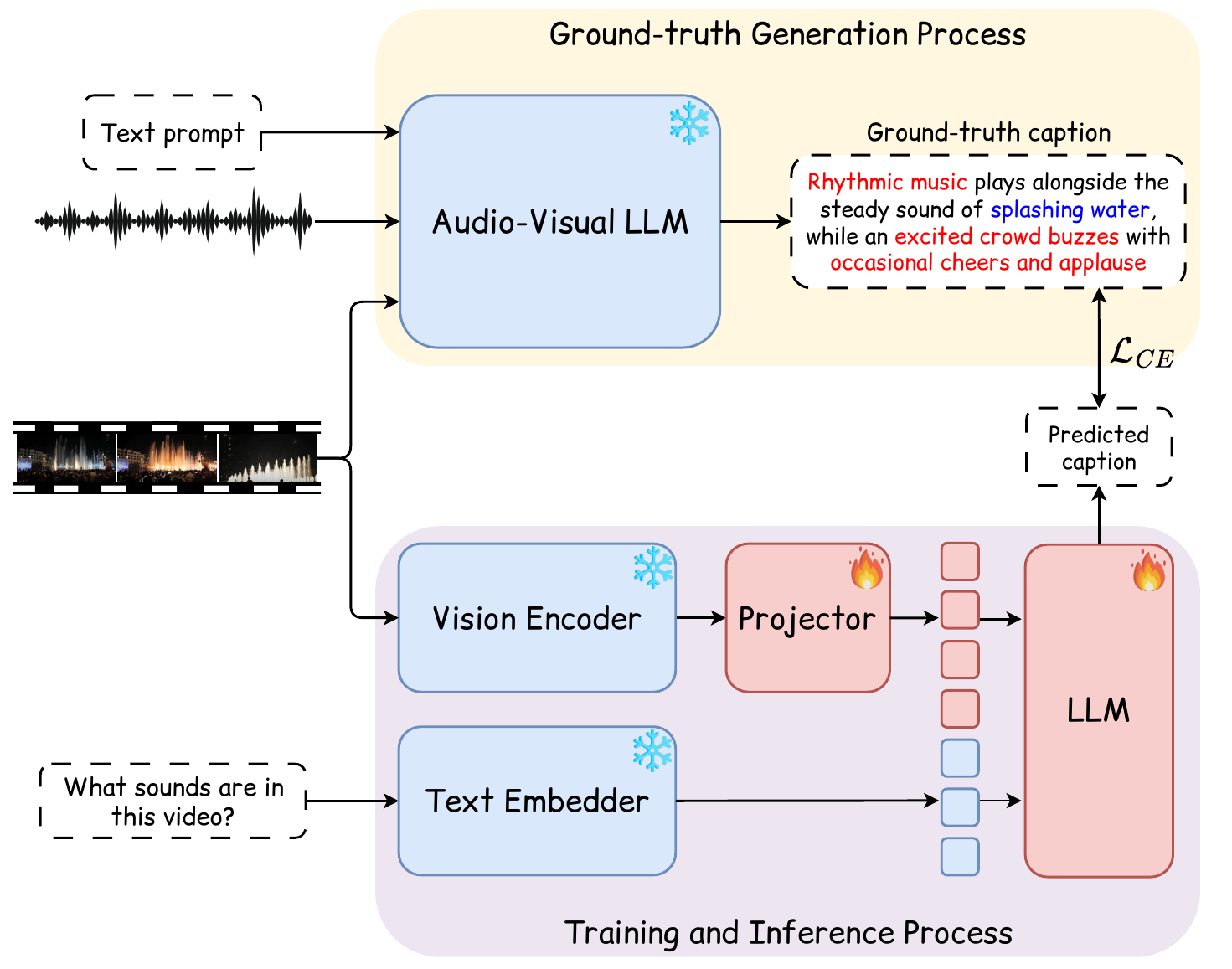}
    \caption{Overview of SoundCap’s ground-truth generation, training, and inference pipeline. Blue labels denote audio events that a standard VLM can detect, while red labels highlight events that are challenging to infer from visuals alone.}
    \label{fig:soundcap}
\end{figure}

Having detailed the network design of Flowley, we next address a complementary challenge: data expressiveness. Existing V2A datasets provide only short labels (most under 6-10 words), limiting the semantic granularity available to the model. SoundCap remedies this limitation by: (1) leveraging an AV‑LLM to craft fine-grained, sound‑oriented captions and (2) employing these captions to supervise a VLM that in turn enhances Flowley’s conditioning signals. As shown in \Cref{fig:soundcap}, for each video-audio pair $(V,A)$ and instruction prompt $T_p = \{ t_p^1,\ldots,t_p^K \}$, the AV‑LLM $P_\Theta(T_a|T_p,V,A)$ produces a detailed audio description $T_a = \{ t_a^1,\ldots,t_a^N \}$. This generated description not only captures both on- and off-screen sounds, but also remains consistent with the video’s visual and acoustic cues, owing to the unified strength of the multimodal model. Next, we treat $T_a$ as ground truth to fine-tune a VLM: feeding it the visual input and a prompt $T_i$, we train it to reproduce the sound-oriented description by minimizing the negative log-likelihood of the target tokens conditioned on the video and instruction:
\begin{equation*}
    \mathcal L_{\text{vlm}}(\hat T_a|T_i, V) = -\sum_{l=1}^N \log[P_\Phi(\hat t_a^l = t_a^l | T_i, V)],
\end{equation*}
where $\Phi$ is the set of trainable weights of VLM. At inference time, only the VLM is executed, enabling sound‑oriented captioning without access to audio. This design explicitly teaches the VLM to recover auditory events that are hard to tell from visuals alone (red) while retaining its ability to describe visually evident (blue) cues (\Cref{fig:soundcap}), thus enriching the conditioning signal later employed by Flowley. Furthermore, as VGGSound is an in-the-wild dataset characterized by inherent noise (e.g., irrelevant background audio or speech), we insert specific instructional ``warnings'' into the prompts to guide the AV-LLM generation. This strategy prevents the VLM from learning from misaligned samples, ensuring more accurate captioning. Further details of the text prompts and examples of audio captions are described in Supplementary Material.

\section{Experimental Results}

\subsection{Experimental Setup}

\subsubsection{Implementation Details.}

Flowley employs $N_1=5$ multi-stream blocks followed by $N_2=10$ single-stream blocks, with the hidden dimension $D=512$. Each Flowley block uses 8 attention heads and a feed-forward network multiplier of four. We apply a dropout rate of 0.1 and limit input text sequences to a maximum length of 77 tokens. We use $\lambda=0.5$ and $\omega=0$ for all experiments, meaning that only exact video-audio frames pair are allowed to fully attend to each other, while they can attend to other frames in fade zone within the range $\delta=4$. At inference, we integrate the ODE using Euler’s method with 25 steps and apply CFG with a strength of $s=7.5$.

For SoundCap training, we use video-SALMONN \cite{pmlr-v235-sun24l} to generate ground-truth text, which is subsequently employed to finetune Qwen2.5-VL \cite{bai2025qwen2} for one epoch. We further emphasize that SoundCap is independent of Flowley and serves as an \textit{optional}, one-time recaptioning step to enrich the captions. Consequently, it is not considered an additional stage in Flowley’s training pipeline. Additional details are provided in the Supplementary Material.

\subsubsection{Datasets and Baselines.}

For a fair comparison with existing baselines, we train Flowley on VGGSound, a standard and large-scale dataset containing approximately 200k 10-second video clips, accompanied by corresponding audio tracks. Following previous baselines, we only use the first 8 seconds of each video for training and evaluation. To provide a comprehensive performance analysis, we benchmark Flowley against seven recent SOTA methods under same standardized conditions, all of which are described in the Supplementary Material.

\subsubsection{Evaluation Metrics.}

We follow previous works to evaluate our method along four dimensions. First, \textbf{distribution matching} measures how closely the statistical properties of generated audio align with those of real recordings. To this end, we calculate the Fréchet Audio Distance (FAD) \cite{10.1109/TASLP.2020.3030497} and Kernel Audio Distance (KAD) \cite{chung2025kad}, both using PANNs \cite{10.1109/TASLP.2020.3030497} as the embedding network, and employ PaSST \cite{koutini22_interspeech} to compute KL divergence in accordance with AudioLDM. Second, \textbf{audio quality} is quantified via the Inception Score (IS) \cite{NIPS2016_8a3363ab}, also leveraging PANNs as the classifier per Frieren. Third, \textbf{semantic alignment} assesses the consistency between the input video content and the synthesized audio by extracting visual and audio embeddings with ImageBind \cite{Girdhar_2023_CVPR} and LanguageBind \cite{zhu2024languagebind} and reporting their average cosine similarity, denoted as the IB-Score and LB-Score, respectively. Finally, \textbf{sync.} evaluates temporal coherence by adopting Alignment Accuracy (Align Acc) metric \cite{NEURIPS2023_98c50f47}.

\subsection{Results}

\begin{table*}[t]
    \setlength{\tabcolsep}{4.5pt}
    \centering
    \caption{Results on the VGGSound test split. Reported parameter counts omit feature encoders, decoders, and vocoders. The $[\dagger]$ denotes results we reproduced using the authors’ official checkpoints and inference scripts; $[\ddagger]$ indicates evaluation on generated samples obtained from the authors; and $[\diamond]$ marks models that were retrained on the VGGSound dataset using the official implementation. \textbf{Best results} are written in bold, and the \underline{runner-up} scores are underlined. Tiny green numbers indicate the {\color{mygreen}performance boost} from incorporating SoundCap.}\label{tab:main-results}
    \resizebox{\textwidth}{!}{
    \begin{tabular}{l|lrrrrrrr}
         \toprule
         \multirow{3}{*}{Method} & \multirow{3}{*}{Params} & \multicolumn{3}{c}{Distribution Matching} & Quality & \multicolumn{2}{c}{Semantic Alignment$(\times 100)$} & Sync. \\
         \cmidrule(lr){3-5} \cmidrule(lr){6-6} \cmidrule(lr){7-8} \cmidrule(lr){9-9}
         & & KAD$\downarrow$ & FAD$\downarrow$ & KL$\downarrow$ & IS$\uparrow$ & IB-Score$\uparrow$ & LB-Score$\uparrow$ & Align Acc$\uparrow$ \\
         \midrule
         Frieren \cite{NEURIPS2024_e7384de3}$^\dagger$ & 159M & 1.27 & 12.8 & 2.82 & 12.02 & 22.45 & 19.09 & \textbf{97.13} \\
         FoleyCrafter \cite{zhang2024foleycrafterbringsilentvideos}$^\dagger$ & 1.22B & 1.54 & 19.17 & 2.19 & 15.09 & 25.75 & {24.66} & 77.15 \\
         V2A-Mapper \cite{Wang_Ma_Pascual_Cartwright_Cai_2024}$^\ddagger$ & 229M & 1.34 & 11.73 & 2.50 & 12.43 & 22.38 & 22.32 & 79.08 \\
         MDSGen \cite{pham2025mdsgen}$^\dagger$ & 131M & 5.33 & 39.68 & 2.85 & 6.87 & 17.75 & 19.05 & \underline{91.70} \\
         Mel-QCD \cite{Wang_2025_CVPR}$^\dagger$ & 859M & 1.53 & 19.17 & {2.09} & 10.32 & 23.79 & 23.80 & 73.85 \\
         VinTAGe \cite{Kushwaha_2025_CVPR}$^\dagger$ & 1.32B & 1.08 & 17.88 & 2.15 & {17.34} & 21.10 & 21.51 & 67.11 \\
         \midrule
         MMAudio \cite{Cheng_2025_CVPR}$^\diamond$ & 157M & 0.57 & 7.89 & 1.91 & 12.68 & 28.09 & 21.98 & 89.73 \\
         $\quad$+ \textbf{SoundCap} & 157M & \tinygreenup{31.6} \textbf{0.39} & \tinygreenup{10.1} \textbf{7.09} & \tinygreenup{18.3} \textbf{1.56} & \tinygreenup{15.8} 14.68 & \tinygreenup{2.7} 28.85 & \tinygreenup{3.0} 22.64 & \tinygreenup{0.8} 90.53 \\
         \midrule
         \textbf{Flowley} & 169M & \underline{0.42} & {7.65} & \underline{1.57} & \underline{18.25} & \underline{29.32} & \underline{24.87} & 89.37 \\
         $\quad$+ \textbf{SoundCap} & 169M & \tinygreenup{7.14} \textbf{0.39} & \tinygreenup{1.7} \underline{7.52} & \tinygreenup{0.6} \textbf{1.56} & \tinygreenup{7.8} \textbf{19.68} & \tinygreenup{2.6} \textbf{30.07} & \tinygreenup{1.8} \textbf{25.33} & \tinygreenup{0.7} 90.02 \\
         \bottomrule
    \end{tabular}}
\end{table*}

\subsubsection{Objective Evaluation.}

As shown in \Cref{tab:main-results}, Flowley outperforms existing baselines across all metrics, with the sole exception of the Align Acc metric. Regarding distribution matching, it leads all methods by significant margins. Specifically, its KAD score of 0.42 represents a 26.3\% gain over the next best baseline, MMAudio (0.57) and nearly triples the performance of the third-ranked method, VinTAGe (1.08). In terms of generation quality, Flowley attains an Inception Score of 18.25, exceeding VinTAGe despite the latter having roughly eight times more parameters. Our synthesized audio further demonstrates exceptional audio-video correspondence, achieving top results on semantic alignment benchmarks. Finally, on synchronization metric, Flowley surpasses approaches that rely on explicit time-conditioned modules (e.g., onset and energy predictors) such as Mel‑QCD and FoleyCrafter. It remains competitive with MMAudio and MDSGen, while only trailing Frieren, all of which utilize a pretrained audio-visual alignment encoder. Notably, \textit{MDSGen and Frieren employ the same pretrained encoder for video feature extraction that serves as the backbone for the Align Acc metric}. This overlap may introduce evaluation bias, potentially inflating the scores of these specific models. To provide a more comprehensive assessment, we conduct a subjective evaluation, detailed in the following section.
 
The four bottom rows of \Cref{tab:main-results} present the results and performance gains obtained by incorporating SoundCap-generated detailed text descriptions into the training and inference pipelines of MMAudio and Flowley. SoundCap acts as a powerful performance multiplier, most notably for MMAudio in the distribution matching category. By incorporating these rich descriptions, MMAudio achieves the best FAD score of 7.09. Furthermore, SoundCap yields a significant 31.6\% improvement in KAD for MMAudio. For Flowley, SoundCap also improves overall performance across several metrics, yielding gains of up to 7.8\%. These findings demonstrate the effectiveness of SoundCap across different frameworks and highlight the importance of rich text descriptions, which are largely absent in current datasets and can be addressed through SoundCap. We provide qualititative examples and analyses in the Supplementary Material.

\subsubsection{Subjective Evaluation.} For a more comprehensive assessment, we complemented our quantitative findings with subjective human evaluations. To mitigate the inherent noise of the VGGSound test set, we rigorously filtered the data by excluding samples with low ImageBind alignment scores, low resolution ($\leq$360p), or human speech. From this subset, we randomly selected 30 videos to construct 180 distinct A/B comparison pairs (Flowley vs. each baseline). Twenty participants were recruited, each rating 36 pairs to ensure every pair was independently evaluated four times, yielding 720 total responses. 

\begin{figure}[t]
    \centering

    \begin{subfigure}[b]{0.325\textwidth}
        \includegraphics[width=\textwidth]{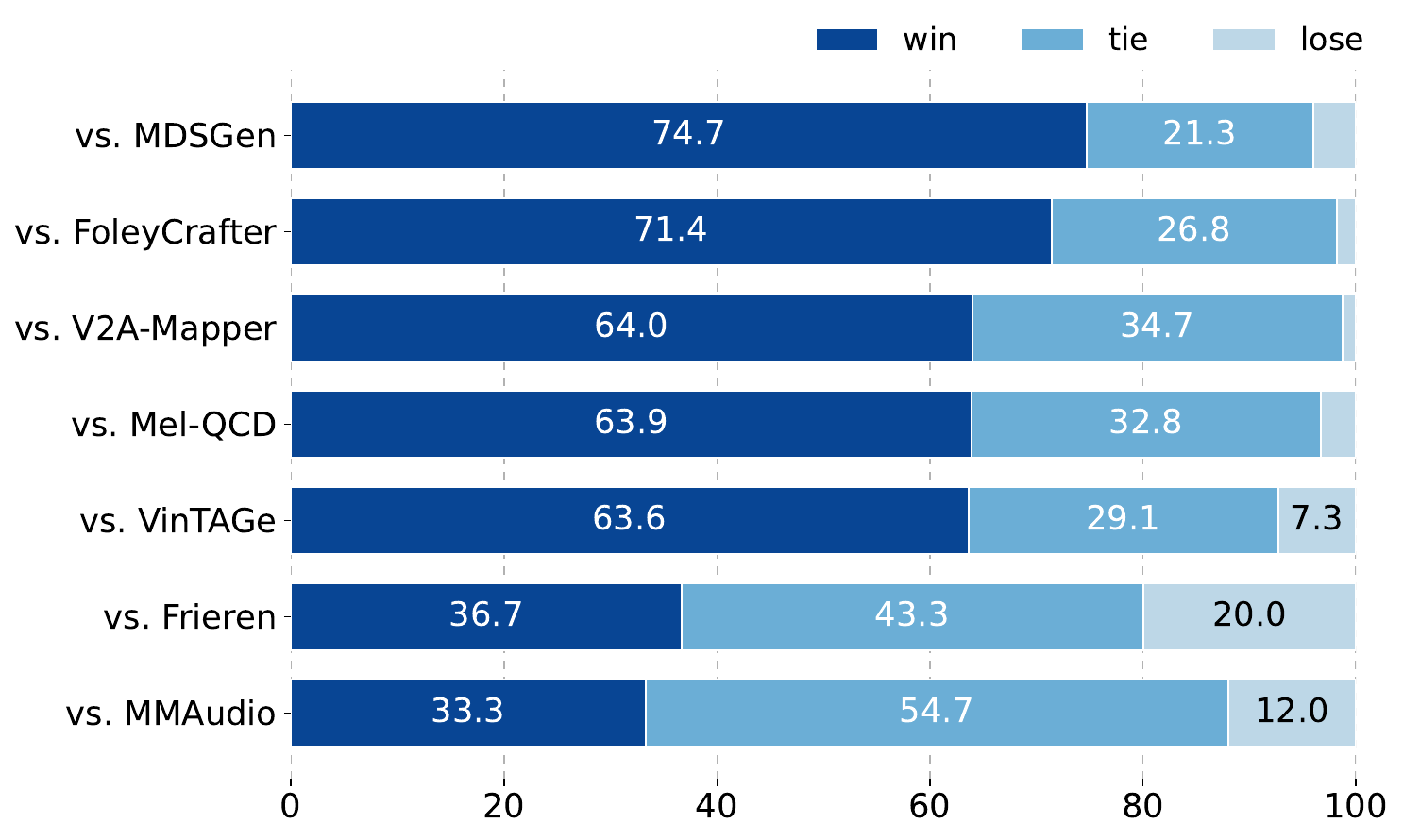}
        \caption{Audio Quality.}
        \label{fig:audio_quality_subjective}
    \end{subfigure}
    \hfill
    \begin{subfigure}[b]{0.325\textwidth}
        \includegraphics[width=\linewidth]{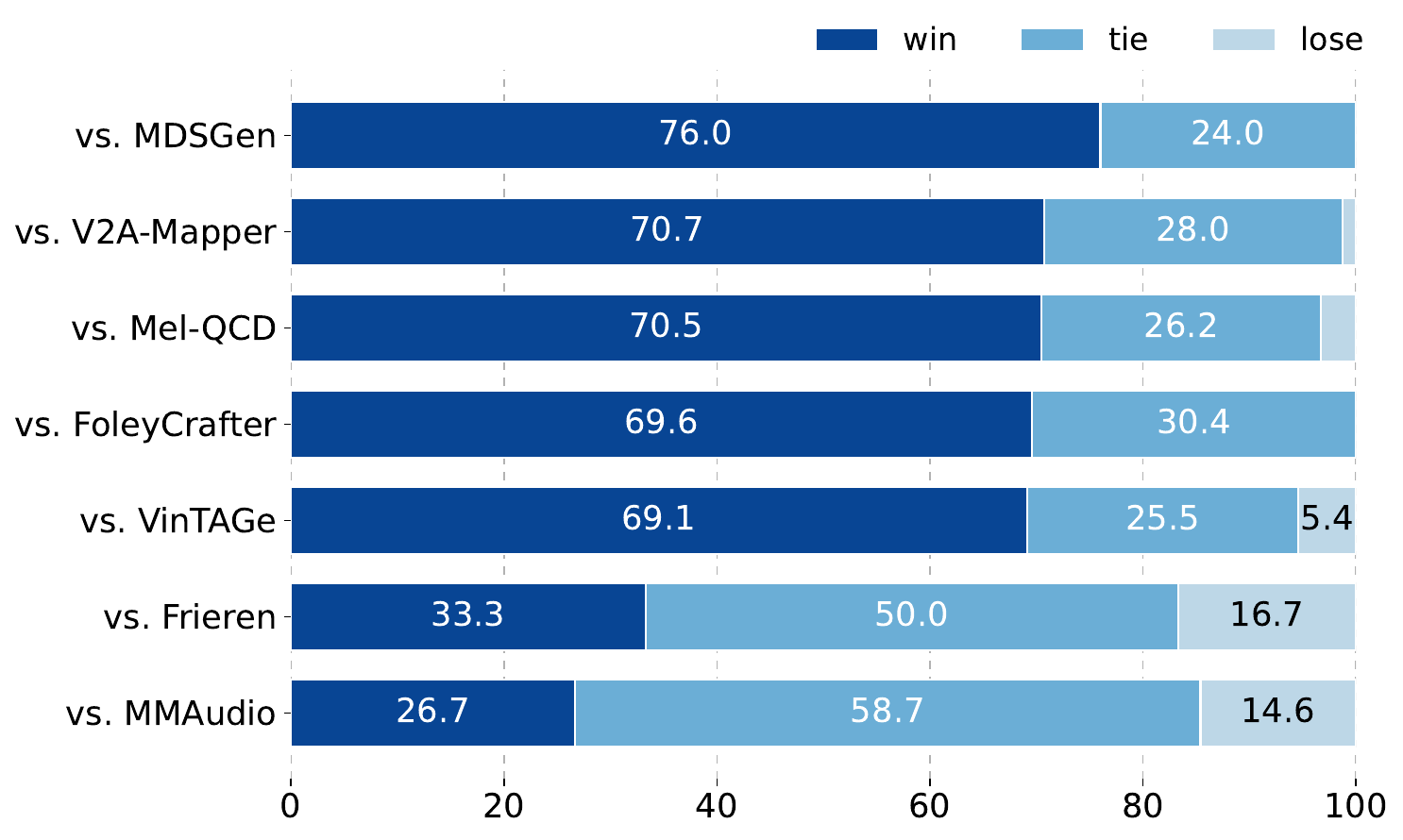}
        \caption{Semantic Alignment.}
        \label{fig:semantic_alignment_subjective}
    \end{subfigure}
    \hfill
    \begin{subfigure}[b]{0.325\textwidth}
        \includegraphics[width=\linewidth]{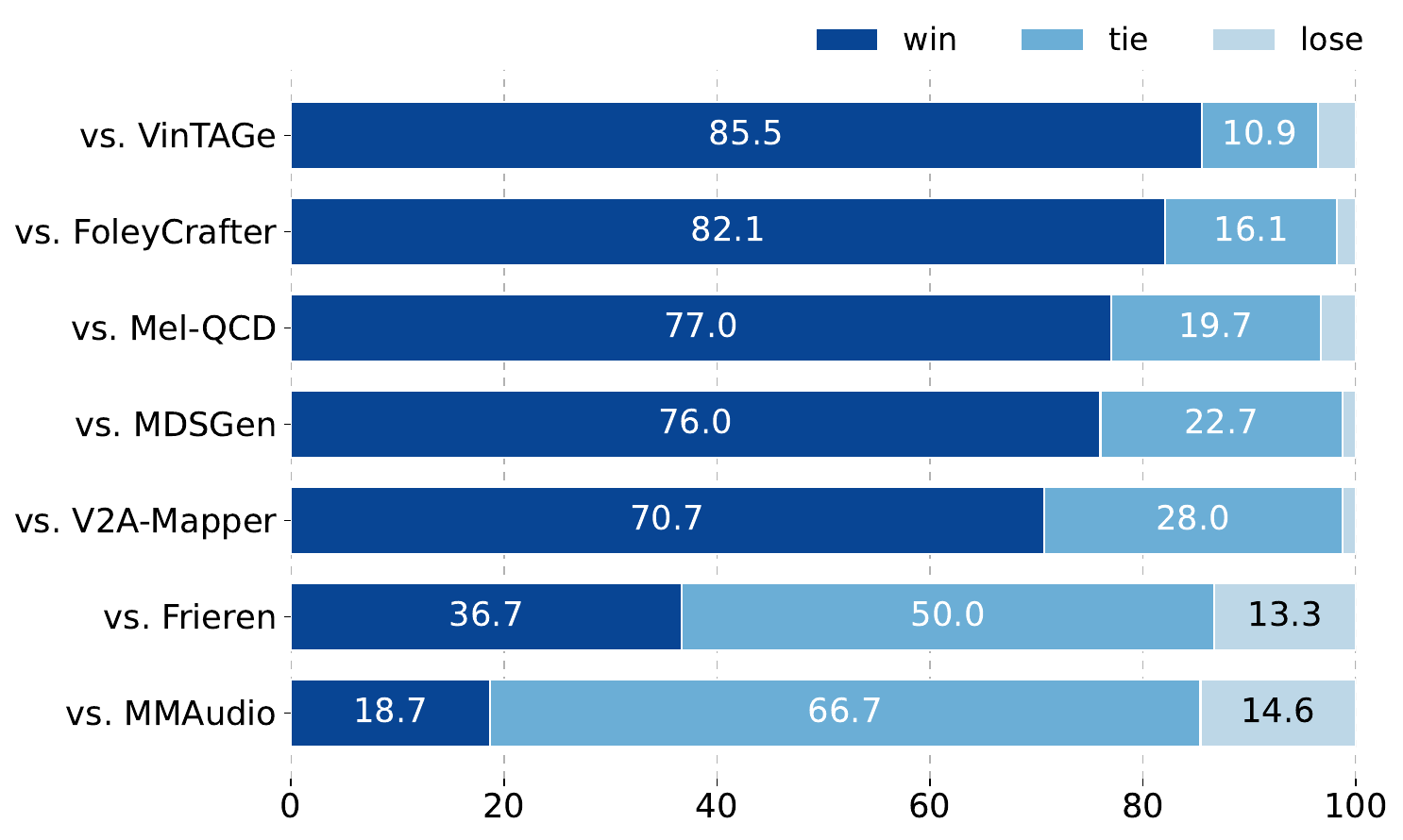}
        \caption{Temporal Alignment.}
        \label{fig:temporal_alignment_subjective}
    \end{subfigure}

    \caption{Human preference comparison between Flowley and competing methods.}
    \label{fig:subjective}
\end{figure}

As illustrated in \Cref{fig:subjective}, Flowley demonstrates a clear and consistent advantage across all categories. Notably, although Flowley trailed MMAudio, MDSGen, and Frieren in the Align Acc metric (\Cref{tab:main-results}), it was preferred over all three in human subjective testing for temporal synchronization. This preference is particularly pronounced against MDSGen, where Flowley achieved a 76\% win rate. These results highlight the effectiveness of our approach and suggest that model-based metrics may not fully capture the perceptual nuances of audio-visual alignment.

\subsection{Ablation Studies}

\subsubsection{Sensitivity of PSCA.}

We evaluate the sensitivity of PSCA and its influence on overall system performance by varying the hard-attention window size $\omega$ and soft-attention zone size $\delta$, as shown in \Cref{tab:psca-sensitivity}. Increasing $\omega$ and $\delta$ generally leads to lower Align Acc metrics, which can be attributed to the expansion of the context window that introduces a greater risk of audio-video misalignment. In contrast, IS and IB-Score remain relatively stable across these hyperparameter settings, suggesting that PSCA maintains robustness in terms of audio quality and semantic consistency. These findings are consistent with our design intuition that PSCA primarily affects audio-video synchronization while leaving overall semantic alignment largely unchanged.

\subsubsection{Modules in Single-stream Blocks.}

We investigate the impact of different configurations of text-audio and visual-audio cross-attention (CA), as well as the effectiveness of the proposed PSCA mechanism. As presented in \Cref{tab:single-abl}, we conduct an ablation study by independently removing the cross-attention to either the textual or visual features (or both) and by replacing the PSCA mechanism with a standard cross-attention module. Our key findings are as follows: (1) Reducing the single-stream block to a vanilla DiT by removing both CA streams results in a significant performance drop across all metrics, highlighting the importance of dual cross-attention (Row 1 \textit{vs.} 6); (2) Incorporating textual features consistently improves the semantic relevance of the generated audio (Rows 2, 5, 6); and (3) Employing the progressive soft mask $M$ improves the temporal synchronization (Rows 4 and 6).

\begin{table}[t]
    \centering
    \begin{minipage}[t]{0.44\textwidth}
        \centering
        \caption{Results obtained by varying the hyperparameters of the PSCA module. Text in blue indicates our \colorbox{lightblue}{default settings}.}
        \label{tab:psca-sensitivity}
        \setlength{\tabcolsep}{6pt}
        \resizebox{0.95\textwidth}{!}{%
        \begin{tabular}{cc|rrr}
            \toprule
            $\omega$ & $\delta$ & IS$\uparrow$ & IB-Score$\uparrow$ & Align Acc$\uparrow$ \\
            \midrule
            \multirow{4}{*}{0} & 0 & 17.85 & 28.84 & 88.56 \\
            & 2 & 18.18 & 29.21 & 88.37 \\
            & \cellcolor{lightblue}4 & \cellcolor{lightblue} 18.25 & \cellcolor{lightblue} 29.32 & \cellcolor{lightblue} \textbf{89.37} \\
            & 8 & 18.27 & 29.34 & 88.02 \\
            \midrule
            \multirow{4}{*}{2} & 0 & 18.15 & \textbf{29.48} & 88.46 \\
            & 2 & \textbf{18.57} & 29.35 & 88.03 \\
            & 4 & 18.2 & 29.44 & 87.58 \\
            & 8 & 18.22 & 29.21 & 87.23 \\
            \midrule
            \multirow{4}{*}{4} & 0 & 18.33 & 29.39 & 88.37 \\
            & 2 & 18.54 & 29.26 & 88.11 \\
            & 4 & 18.48 & 29.36 & 87.22 \\
            & 8 & 18.52 & 29.18 & 86.58 \\
            \bottomrule
        \end{tabular}%
        }
    \end{minipage}\hfill
    \begin{minipage}[t]{0.54\textwidth}
        \centering
        
        \caption{Results obtained by altering single-modal stream block.\label{tab:single-abl}}
        \setlength{\tabcolsep}{3pt}
        \resizebox{0.95\textwidth}{!}{%
        \begin{tabular}{c|ccc|rrrr}
            \toprule
            \# & CA$_t$ & CA$_v$ & PSCA$_v$ & KAD$\downarrow$ & IS$\uparrow$ & IB-Score$\uparrow$ & Align Acc$\uparrow$ \\
            \midrule
            1 & \xmark & \xmark & \xmark & {0.44} & 16.99 & 27.38 & 86.26 \\
            2 & \cmark & \xmark & \xmark & {0.44} & 17.29 & \underline{28.93} & 87.43 \\
            3 & \xmark & \cmark & \xmark & \textbf{0.40} & 17.58 & 28.02 & 87.04 \\
            4 & \xmark & \xmark & \cmark & \textbf{0.40} & 17.69 & 28.51 & \underline{88.72} \\
            5 & \cmark & \cmark & \xmark & \textbf{0.40} & \underline{18.15} & 28.74 & 87.61 \\
            \midrule
            6 & \cmark & \xmark & \cmark & \underline{0.42} & \textbf{18.25} & \textbf{29.32} & \textbf{89.37} \\
            \bottomrule
        \end{tabular}%
        }
        
        \vspace{0.7cm} 
        
        \caption{Results obtained by removing layer-dependent progressive parameter.}
        \label{tab:abla-beta}
        \setlength{\tabcolsep}{6pt}
        \resizebox{0.9\textwidth}{!}{%
        \begin{tabular}{l|rrrr}
            \toprule
            $\beta$ & KAD$\downarrow$ & IS$\uparrow$ & IB-Score$\uparrow$ & Align Acc$\uparrow$ \\
            \midrule
            \xmark & \textbf{0.42} & 17.78 & \textbf{29.53} & 88.62 \\
            \cmark & \textbf{0.42} & \textbf{18.25} & 29.32 & \textbf{89.37} \\
            \bottomrule
        \end{tabular}%
        }
    \end{minipage}
\end{table}

\begin{table}[t]
    \centering
    \caption{Ablation study on the impact of noise-aware conditioning in SoundCap.}
    \label{tab:abla-soundcap-noise}
    \setlength{\tabcolsep}{5pt}
    \begin{tabular}{l|rrrr}
        \toprule
        Method & KAD$\downarrow$ & IS$\uparrow$ & IB-Score$\uparrow$ & Align Acc$\uparrow$ \\
        \midrule
        Flowley & \underline{0.42} & \underline{18.25} & 29.32 & \underline{89.37} \\
        \quad + SoundCap w/o noise conditions & 0.45 & 15.16 & \underline{29.61} & 87.13 \\
        \quad + SoundCap w/ noise conditions & \textbf{0.39} & \textbf{19.68} & \textbf{30.07} & \textbf{90.02} \\
        \bottomrule
    \end{tabular}
\end{table}

\subsubsection{Impact of layer-dependent progressive parameter.} To understand the contribution of $\beta$ to the overall performance, we perform the ablation study presented in \Cref{tab:abla-beta}, where the depth-aware parameter is removed from the cross-attention masking calculation. The results show that this omission leads to a notable decline in both IS and Align Acc. While the IB-Score remains relatively stable, the degradation in audio quality and temporal synchronization metrics underscores the critical role of $\beta$ in achieving precise audio-visual alignment.

\subsubsection{Handling noise in SoundCap.} Following the discussion in \Cref{sec:soundcap}, we mitigate the inherent noise of the VGGSound dataset by inserting specific noise conditions into the text prompts. The effectiveness of this approach is evaluated in \Cref{tab:abla-soundcap-noise}, where we compare Flowley's performance using two distinct prompt configurations (detailed in the Supplementary Material). The results demonstrate that our noise-handling strategy yields consistent improvements across all metrics. Conversely, when SoundCap is unaware of these conditions, audio quality significantly degrades: IS drops by 16.9\%, Align Acc decreases to 87.13\%, and KAD rises marginally. These findings underscore that noise-robust conditioning is vital for effectively leveraging in-the-wild datasets, as it prevents the model from converging on misaligned or low-quality audio-visual pairs.

\subsubsection{Zero-shot Performance.}

\begin{table}[t]
    \setlength{\tabcolsep}{6pt}
    \centering
    \caption{Zero-shot performance on MovieGen Audio Bench.}\label{tab:moviegen}
    \begin{tabular}{lllrrr}
        \toprule
        Method & Params & Data(h) & IS$\uparrow$ & IB-Score$\uparrow$ & Align Acc$\uparrow$ \\
        \midrule
        Movie Gen Audio & 13B & $\mathcal O(1\text{M})$ & {8.01} & \textbf{35.86} & \textbf{64.33} \\
        \midrule
        Flowley & 169M & $\sim400$ & {7.74} & {23.65} & 59.28 \\
        $\quad$+ SoundCap & 169M & $\sim400$ & \textbf{8.18} & {25.78} & 61.07 \\
        \bottomrule
    \end{tabular}
\end{table}

Finally, we explore the boundaries of our method by testing Flowley on the Movie Gen Audio Bench \cite{polyak2024movie}, a synthetic benchmark dataset whose distribution deviates significantly from that of VGGSound. \Cref{tab:moviegen} summarizes our results. Surprisingly, despite being $77\times$ smaller and trained on just $1/2500$ of the data, Flowley performs comparably to Movie Gen Audio \cite{polyak2024movie} in terms of audio quality (0.27 difference) and outperforms it when SoundCap is incorporated into the inference pipeline, while marginally falling behind in Align Acc by 3.26\%. Regarding semantic alignment, Movie Gen Audio outperforms both versions by a notable margin. We believe this is due to the limited scope of our training data, which restricts our model’s ability to generalize to the wider and more varied content present in the benchmark. We present additional details and results in the Supplementary Material.

\section{Conclusion}

We present Flowley, an end-to-end, flow-based multimodal framework for generating audio that is both semantically meaningful and temporally synchronized with silent video content. Unlike prior works that rely on multi-stage training pipelines or pretrained audio-visual alignment modules, Flowley directly learns both semantic and temporal alignment through the integration of dual cross-attention pathways and a task-specific progressive masking mechanism tailored for video-to-audio synthesis. To further enhance semantic fidelity, we propose SoundCap, a dedicated sound-aware captioning pipeline that provides enriched textual conditioning for guiding audio generation. Empirical evaluations show that Flowley achieves state-of-the-art results in VGGSound across multiple metrics, outperforming models up to $8\times$ larger in parameter size. Moreover, when SoundCap is incorporated into the inference pipeline, Flowley delivers the highest audio quality in a zero-shot setting, outperforming a model $77\times$ larger and trained on $2500\times$ more data on a synthetic evaluation set. Comprehensive ablation studies confirm the effectiveness of each architectural component and highlight the robustness of the overall approach.


%
%
\bibliographystyle{splncs04}
\bibliography{main}

\end{document}


\title{\textit{Supplementary Material for}\\Precise Video-to-Audio Generation with Cross-Modal Alignment in Latent Space} 

\titlerunning{Flowley}

\author{Thanh V. T. Tran\inst{1} \and
Ngoc-Son Nguyen\inst{1} \and
Luong Tran\inst{1} \and
Long-Khanh Pham\inst{1} \and
Paarth Neekhara\inst{2} \and
Shehzeen Hussain\inst{2} \and
Van Nguyen\inst{1}} 

\authorrunning{Tran et al.}

\institute{FPT Software AI Center, Vietnam \and
NVIDIA Corporation, USA \\
\url{https://flowley-v2a.github.io} \\
\email{\{thanhtvt1,sonnn45,luongtk,khanhpl2,vannth19\}@fpt.com}
\email{\{pneekhara,shehzeenh\}@nvidia.com}}

\maketitle

\section{Tangent-based Sampling Schedule Derivation}

This section derives the tangent-based timestep sampling distribution used in our flow matching framework (Section 3.4, main paper). We show that it emerges naturally from the DDPM cosine schedule \cite{improved-ddpm} by preserving the signal-to-noise ratio (SNR) trajectory under the rectified flow interpolation.

\subsection{DDPM Cosine Schedule}

Nichol \& Dhariwal \cite{improved-ddpm} define the cosine noise schedule as:
\begin{equation}\label{eq:cos_noise_sche}
    \overline{\alpha}_t = \cos^2 \bigg(\frac{\pi}{2} \tau \bigg),\quad \tau:=\frac{t}{T} \in [0,1],
\end{equation}
where we take the limit $s \rightarrow 0$ for a clean continuous-time derivation. The DDPM forward process is given by:
\begin{equation}\label{eq:ddpm}
    x_t = \sqrt{\overline{\alpha}_\tau}x_0 + \sqrt{1-\overline{\alpha}_\tau} \epsilon,\quad \epsilon \in \mathcal N(0,\mathbf{I}),
\end{equation}
where $x_0$ is the clean data distribution. Substitute \Cref{eq:cos_noise_sche} into \Cref{eq:ddpm} yields the signal and noise coefficients:
\begin{equation*}
    \sqrt{\overline{\alpha}_\tau} = \cos\bigg(\frac{\pi}{2}\tau\bigg),\quad \sqrt{1-\overline{\alpha}_\tau} = \sin\bigg(\frac{\pi}{2}\tau\bigg).
\end{equation*}

The signal-to-noise ratio (SNR) of the DDPM forward process at time $\tau$ is therefore
\begin{equation}\label{eq:snr-ddpm}
    \text{SNR}_{\text{DDPM}}(\tau) = \frac{\cos\Big(\frac{\pi}{2}\tau\Big)}{\sin\Big(\frac{\pi}{2}\tau\Big)} = \tan\bigg(\frac{\pi}{2} (1-\tau)\bigg).
\end{equation}

\subsection{Flow Matching SNR}

In our flow matching framework, we train the model to predict the velocity of the probability path defined by the linear interpolation $x_t=tx_1+(1-t)x_0$ with $t\in[0,1]$, where $t=0$ is pure Gaussian noise and $t=1$ is clean data. The SNR is simply:
\begin{equation}\label{eq:snr-fm}
    \text{SNR}_{\text{FM}}(t) = \frac{t}{1-t}.
\end{equation}

\subsection{SNR-Preserving Timestep Mapping}

To ensure that the flow matching trajectory traverses the same relative balance of signal and noise as the DDPM cosine schedule, we equate the SNRs of the two processes. Let $u:=1-\tau$; since DDPM training samples $\tau$ uniformly, $u\sim \mathcal U[0,1]$. Equating \Cref{eq:snr-ddpm} and \Cref{eq:snr-fm} yields:
\begin{equation}\label{eq:equate}
    \frac{t}{1-t} = \tan\bigg(\frac{\pi}{2} u\bigg).
\end{equation}

Solving \Cref{eq:equate} for $t$ gives a clean, compact form utilized for sampling in our main paper:
\begin{equation*}
    t=1-\frac{1}{\tan\Big(\frac{\pi}{2} u\Big) + 1}. 
\end{equation*}

\section{Experimental Setup}\label{abla:exp_setup}

\subsection{Implementation Details}\label{abla:implementation_details}

For training, we use a base learning rate of 1.5e-4 with a linear warm-up over the first 1,000 steps, training for a total of 60,000 iterations with a batch size of 512. Optimization is performed using the AdamW optimizer \cite{loshchilov2018decoupled}, with $\beta_1=0.9$, $\beta_2=0.95$, and a weight decay of 1e-6. We employ the post-hoc exponential moving average (EMA) strategy \cite{Karras_2024_CVPR} with a relative width $\sigma_{\text{rel}}=0.05$. To improve training efficiency, we use \texttt{bf16} mixed-precision training. Additionally, all audio latents, visual embeddings, and textual features are precomputed offline and loaded during training. All the Flowley training are conducted on $2\times$ 80GB NVIDIA A100 GPUs.

\begin{figure}[t]
\centering
\begin{minipage}[t]{0.49\linewidth}
  \centering
  \includegraphics[width=\textwidth]{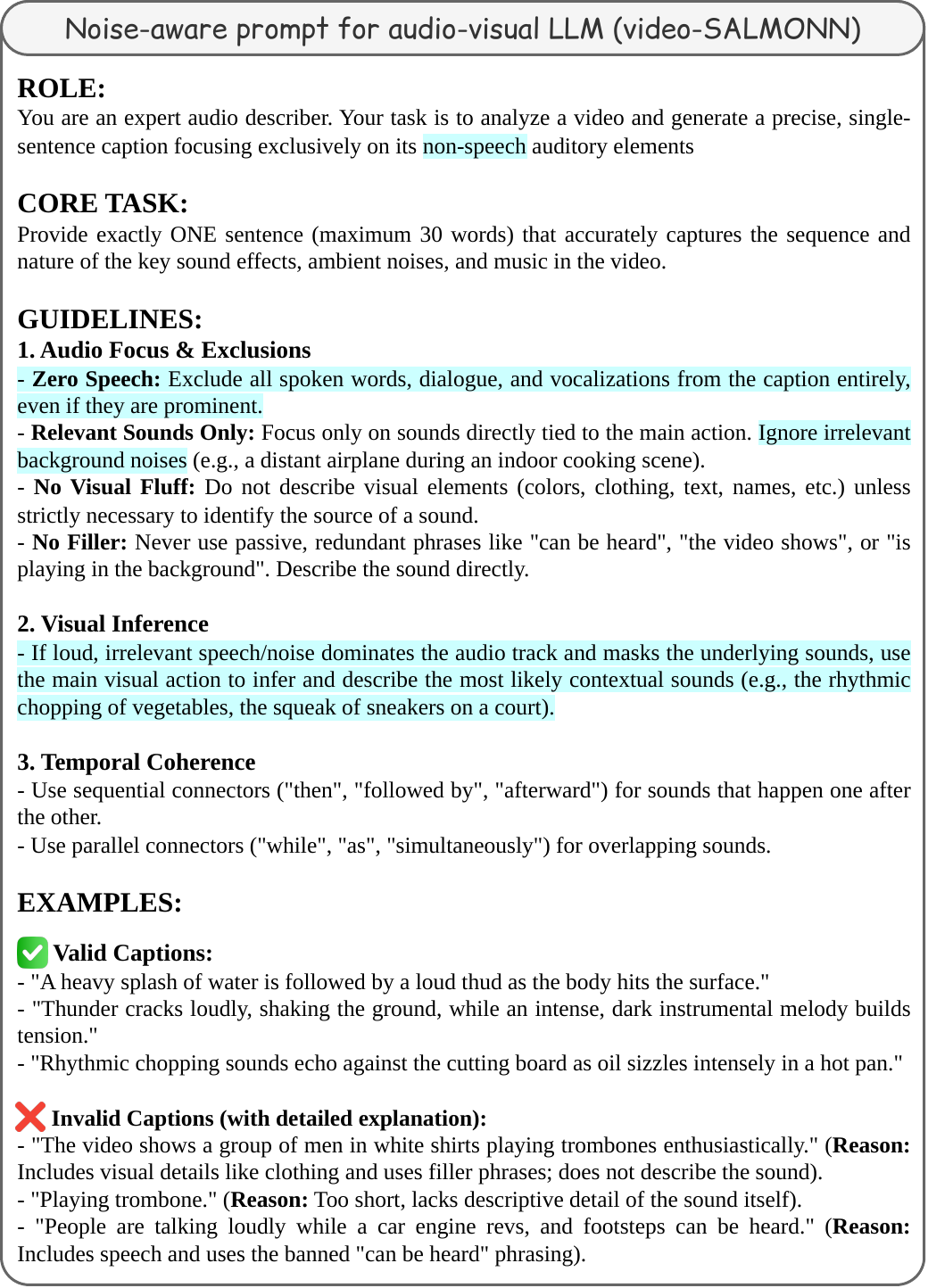}
  \caption{Noise-aware (blue) prompt configuration for robust audio-visual LLM generation.}
  \label{fig:avllm-prompt-conditions}
\end{minipage}%
\hfill
\begin{minipage}[t]{0.49\linewidth}
  \centering
  \includegraphics[width=\textwidth]{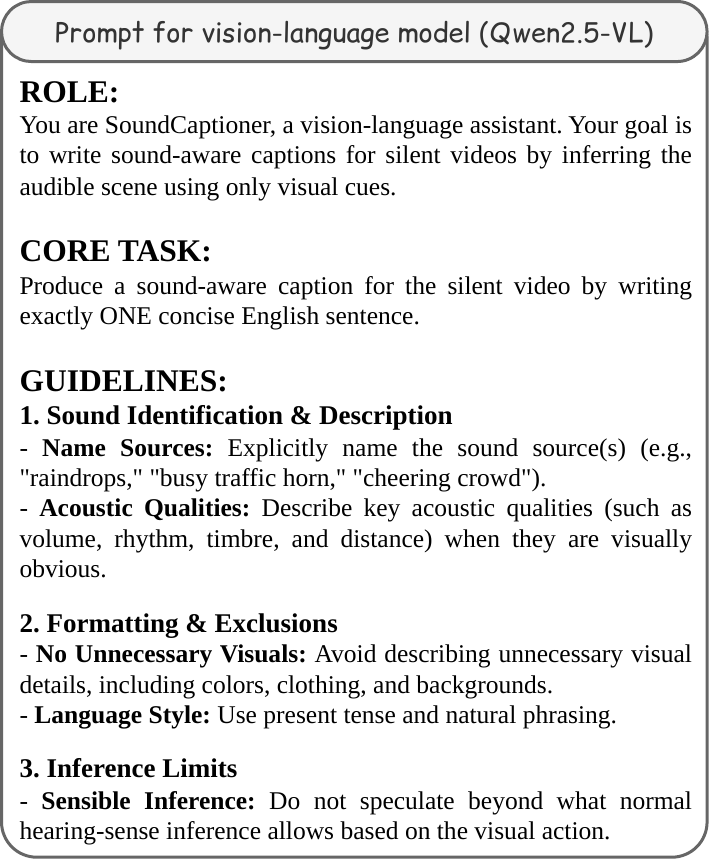}
  \caption{The prompts used for VLMs fine-tuning and synthesizing captions.}
  \label{fig:vlm-prompt}
\end{minipage}
\end{figure}

In addition, we detail below the prompts used in the SoundCap pipeline. It is important to note that VGGSound is an in-the-wild dataset containing many samples whose sound events cannot be reliably inferred from silent video alone (e.g., overlapping speech, irrelevant audio). Although an additional dataset-filtering stage could mitigate this issue, we aim to evaluate the robustness of both the prompts and the models under noisy conditions. Accordingly, we explicitly prompt the AV-LLM to handle such cases. Figure \ref{fig:avllm-prompt-conditions} presents the instruction fed to video-SALMONN \cite{pmlr-v235-sun24l} for generating sound-aware descriptions. Subsequently, we use the prompt shown in Figure \ref{fig:vlm-prompt} together with the silent video to fine-tune Qwen2.5-VL \cite{bai2025qwen2}. The resulting VLM can then produce detailed audio captions, which serve as input prompts for Flowley or any multimodal frameworks. It is worth emphasizing that although our method utilizes these two specific models, practitioners are free to substitute their own AV-LLMs and VLMs within the SoundCap pipeline as modular components.

\begin{figure}
\centering
\begin{minipage}[t]{0.49\linewidth}
  \centering
  \includegraphics[width=\textwidth]{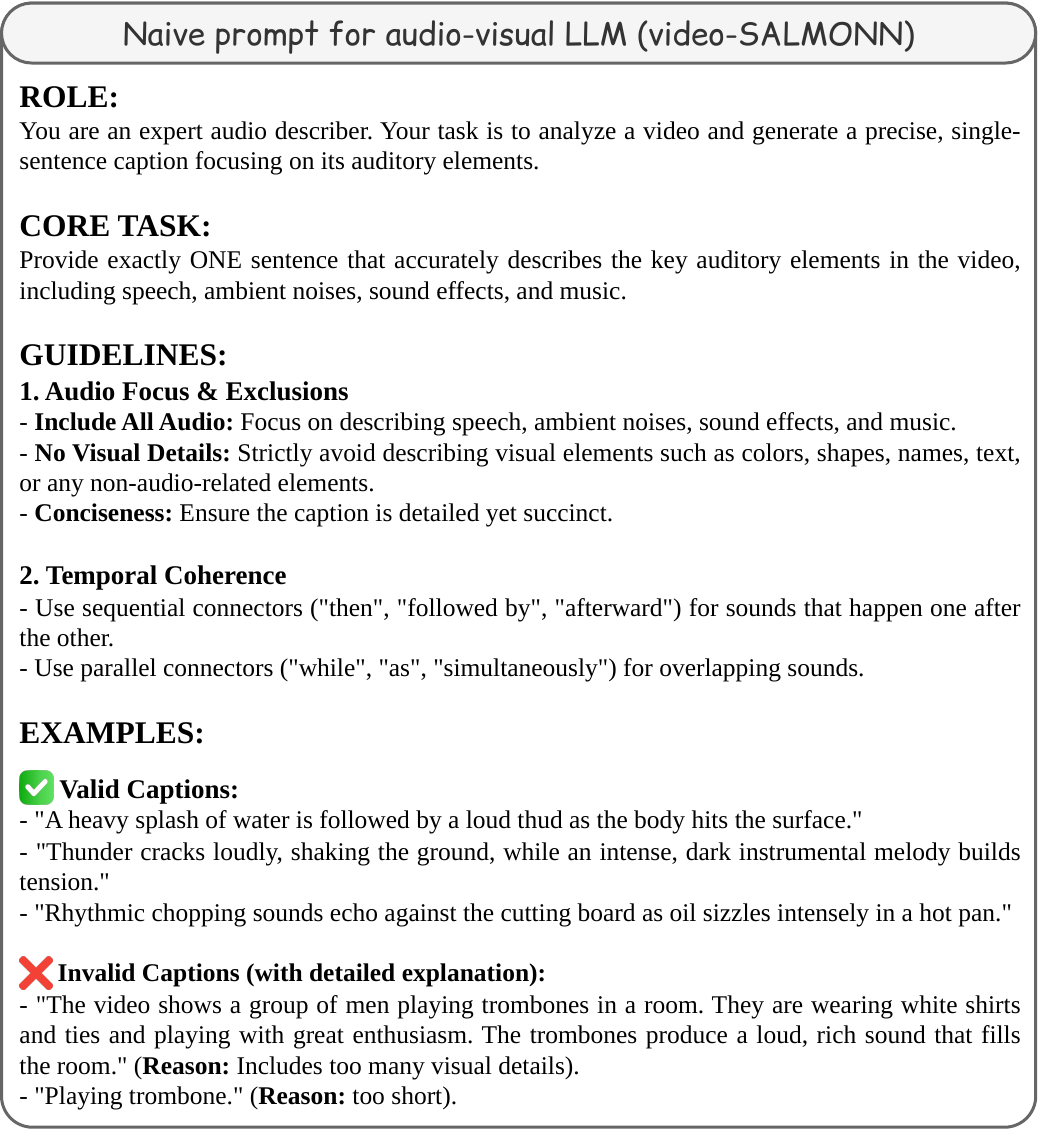}
  \caption{Noise-naive prompt configuration.}
  \label{fig:avllm-prompt-naive}
\end{minipage}%
\hfill
\begin{minipage}[t]{0.49\linewidth}
  \centering
  \includegraphics[width=\textwidth]{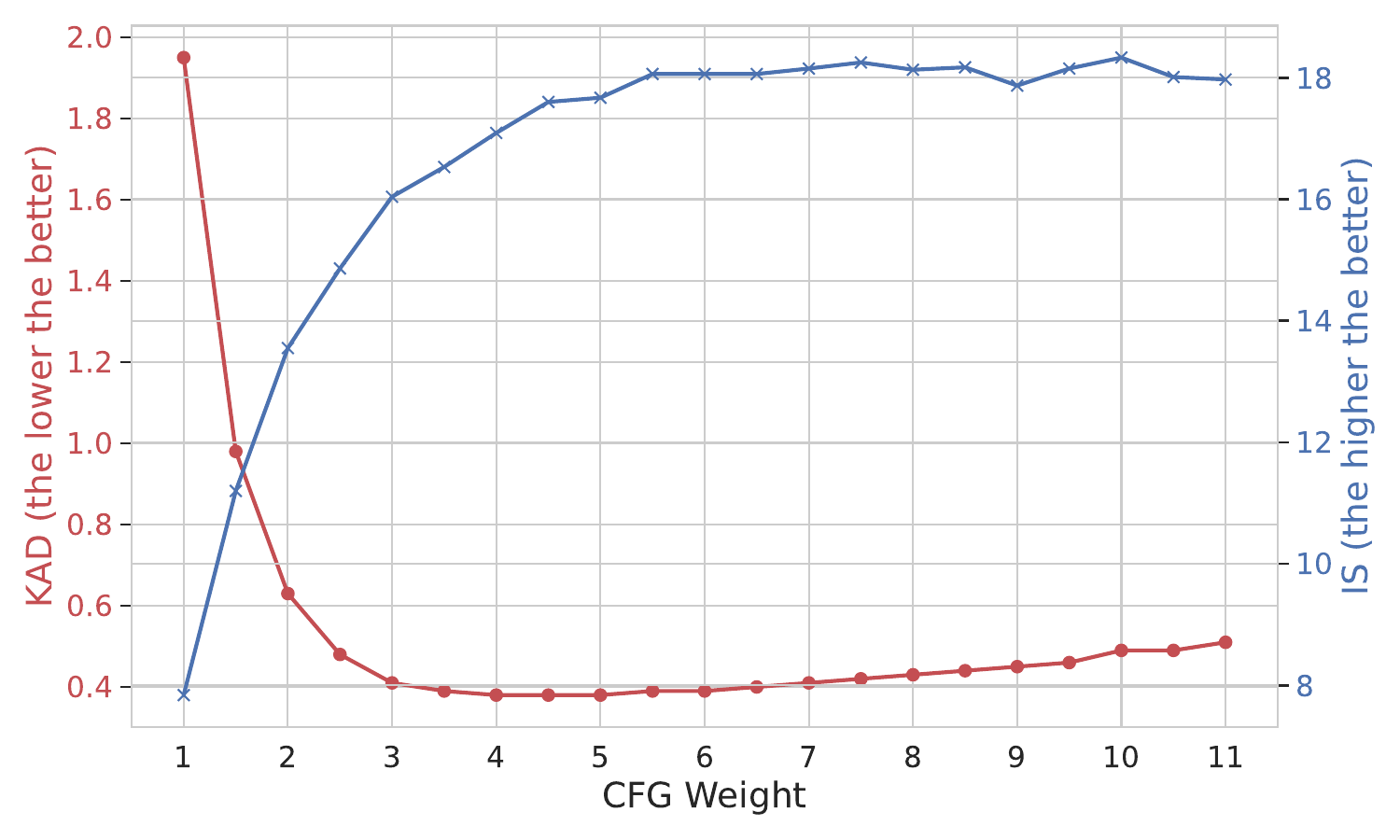}
  \caption{Results when we vary CFG weight.}
  \label{fig:cfg}
\end{minipage}
\end{figure}

\subsection{Baselines}\label{abla:baselines}

We evaluate Flowley against several recent state-of-the-art methods:
\begin{itemize}
    \item Frieren \cite{NEURIPS2024_e7384de3} employs rectified flow matching for V2A synthesis;
    \item FoleyCrafter \cite{zhang2024foleycrafterbringsilentvideos} adapts a pretrained T2A model with semantic and temporal adapters to enforce audio–video alignment;
    \item V2A-Mapper \cite{Wang_Ma_Pascual_Cartwright_Cai_2024} projects visual inputs into text space to condition a T2A generator;
    \item MDSGen \cite{pham2025mdsgen} incorporates a masking strategy to model temporal dependencies in audio;
    \item Mel-QCD \cite{Wang_2025_CVPR} decomposes mel-spectrograms into three signal components and predicts each from video via dedicated predictors;
    \item VinTAGe \cite{Kushwaha_2025_CVPR} is a flow-based transformer that jointly conditions on text and video for audio generation;
    \item MMAudio \cite{Cheng_2025_CVPR} introduces a frame-level conditional synchronization module to enhance audio-visual synchrony.
\end{itemize}
All methods are benchmarked using 25-step Euler or DDIM sampling.

\section{Ablation Study}\label{abla:abla}

\subsection{Handling Noise in SoundCap}

\Cref{fig:avllm-prompt-naive} illustrates the baseline (noise-naive) prompt configuration used in our ablation study to evaluate the impact of noise-aware conditioning on model performance. The primary distinction lies in the inclusion of targeted ``warnings'' in the noise-aware configuration, which explicitly instruct the model to exclude non-essential speech and irrelevant background sounds while utilizing visual inference to recover masked auditory events.

\subsection{CFG Scale}

We present the performance of Flowley under varying CFG scales in \Cref{fig:cfg}. The results indicate that the optimal balance between generation quality and distribution alignment is achieved when $s \in [5.5, 7.5]$. Reducing $s$ below this range leads to a drop in both metrics, while increasing it beyond this point degrades KAD performance. Based on this observation, we set $s=7.5$ for all reported experiments.

\subsection{Zero-shot Performance on Movie Gen Audio Bench}\label{abla:moviegenaudio}

Movie Gen Audio \cite{polyak2024movie} consists of 36 DiT \cite{pmlr-v235-esser24a} layers with attention and feed-forward dimensions of 4,608 and 18,432, respectively, totaling approximately 13B parameters. Although it is not open-sourced, Movie Gen Audio is widely recognized as the industry's state-of-the-art solution for V2A generation. The model is trained on a proprietary dataset that is over 2,500 times larger than VGGSound \cite{9053174}.

Movie Gen Audio Bench \cite{polyak2024movie} is a benchmark dataset comprising 527 video samples generated using Movie Gen Video \cite{polyak2024movie}. As the videos are synthetically generated, the dataset distribution significantly differs from real-world content, often exhibiting characteristics such as slow motion or overly smoothed textures. Moreover, because these videos are artificial, no ground-truth audio is available. Consequently, evaluation is limited to: Inception Score (IS) for audio quality, IB-Score (ImageBind similarity \cite{Girdhar_2023_CVPR}) for semantic alignment between video and audio, and Align Acc (alignment accuracy predicted by CAVP \cite{NEURIPS2023_98c50f47}) for audio-visual synchronization accuracy.

\begin{figure}[t]
    \centering
    \includegraphics[width=\linewidth]{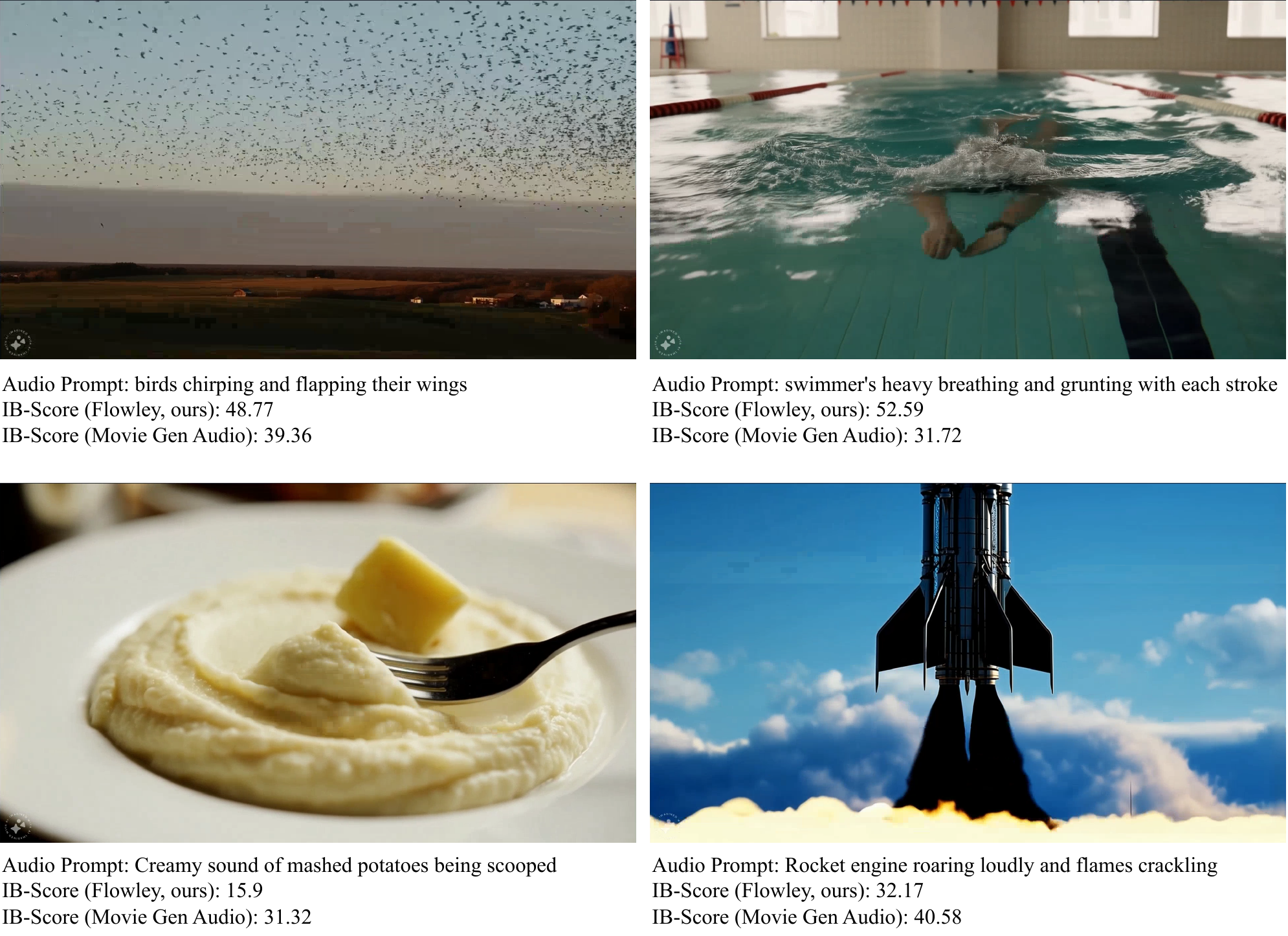}
    \caption{Examples of videos in Movie Gen Audio Bench that are well/not well covered by VGGSound. \textbf{Top:} Flowley achieves higher IB-Scores on videos featuring familiar concepts (birds with 2,508 samples and swimming with 516), which are well represented in the training data. \textbf{Bottom:} Performance drops for unfamiliar concepts (there are no videos related to mashed potatoes or rockets), which are absent from VGGSound, leading to lower IB-Scores.}
    \label{fig:moviegen-results}
\end{figure}

We have summarized our results in the main text (Table 6), showing that our method performs comparably with Movie Gen Audio, with the exception in the IB-Score metric. We believe this is much due to the limited scope in the training data that is not enough to adequately cover the data in Movie Gen Audio Bench and thus falls short in unfamiliar video types. \Cref{fig:moviegen-results} presents examples where Flowley yields notably better or worse IB-Scores, corresponding to whether the video concepts are familiar or rare in the training set.

\section{Limitations}

While our noise-aware prompting significantly mitigates artifacts in VGGSound, raw misalignments can still occasionally surface. Thus, Flowley still faces three primary limitations: (1) unintelligible speech generation; (2) lower-quality background music due to lack of musical training; and (3) out-of-distribution generalization (\Cref{abla:moviegenaudio}). Importantly, because our core objective is precise sound effect synthesis, complex human speech (Limitation 1) is outside our scope. We firmly believe the remaining constraints are fundamentally data-bound and resolvable by scaling with larger, higher-quality curated datasets.

\section{Examples of Synthesized Sound-Oriented Caption}\label{abla:soundcap_examples}

\begin{figure}[t]
    \centering

    \begin{subfigure}[b]{0.49\textwidth}
        \includegraphics[width=\linewidth]{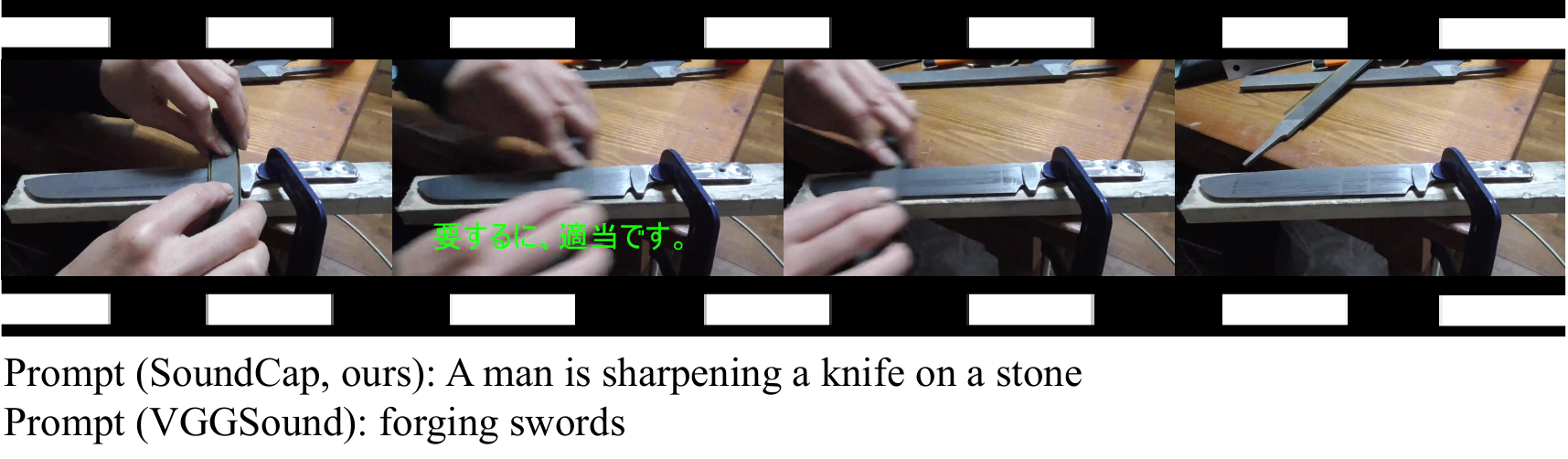}
        \caption{Synthesized caption for video id \texttt{tB67geb0fVg\_000063}.}
        \label{fig:sub1}
    \end{subfigure}
    \hfill
    \begin{subfigure}[b]{0.49\textwidth}
        \includegraphics[width=\linewidth]{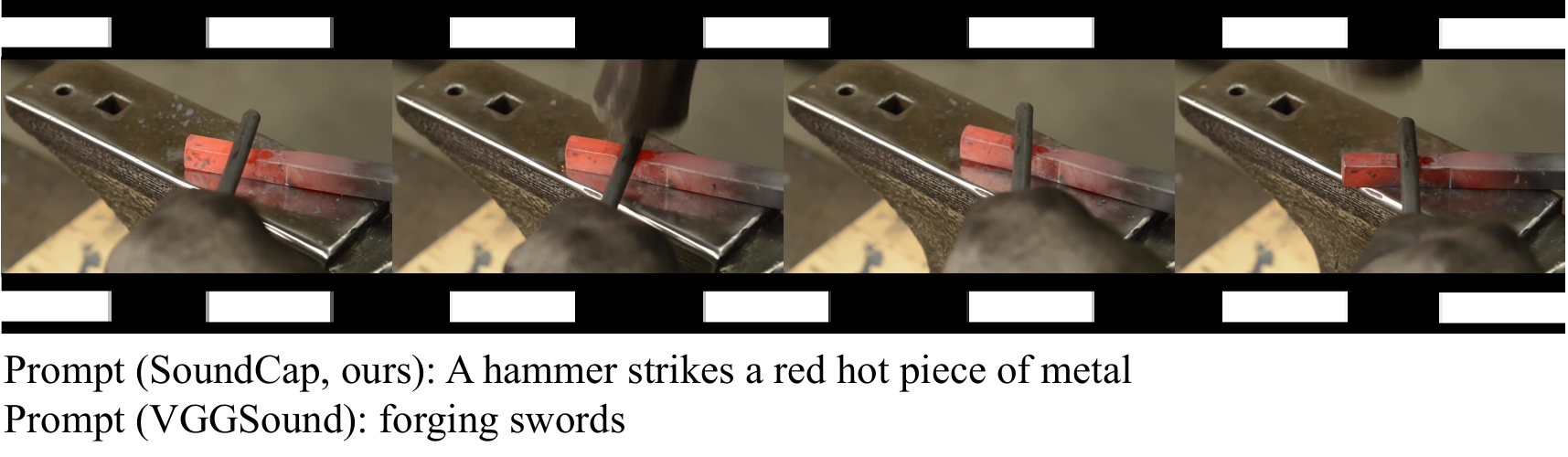}
        \caption{Synthesized caption for video id \texttt{U1IJPJ4kZ-A\_000270}.}
        \label{fig:sub2}
    \end{subfigure}
    \begin{subfigure}[b]{0.49\textwidth}
        \includegraphics[width=\linewidth]{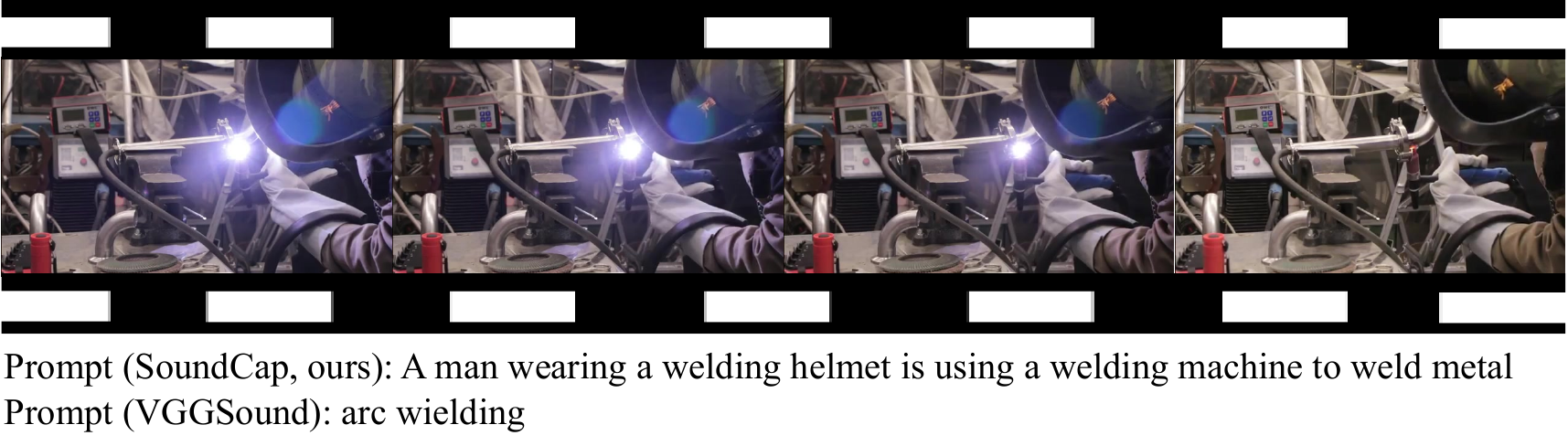}
        \caption{Synthesized caption for video id \texttt{NsUPB1V0c50\_000099}.}
        \label{fig:sub3}
    \end{subfigure}
    \hfill
    \begin{subfigure}[b]{0.49\textwidth}
        \includegraphics[width=\linewidth]{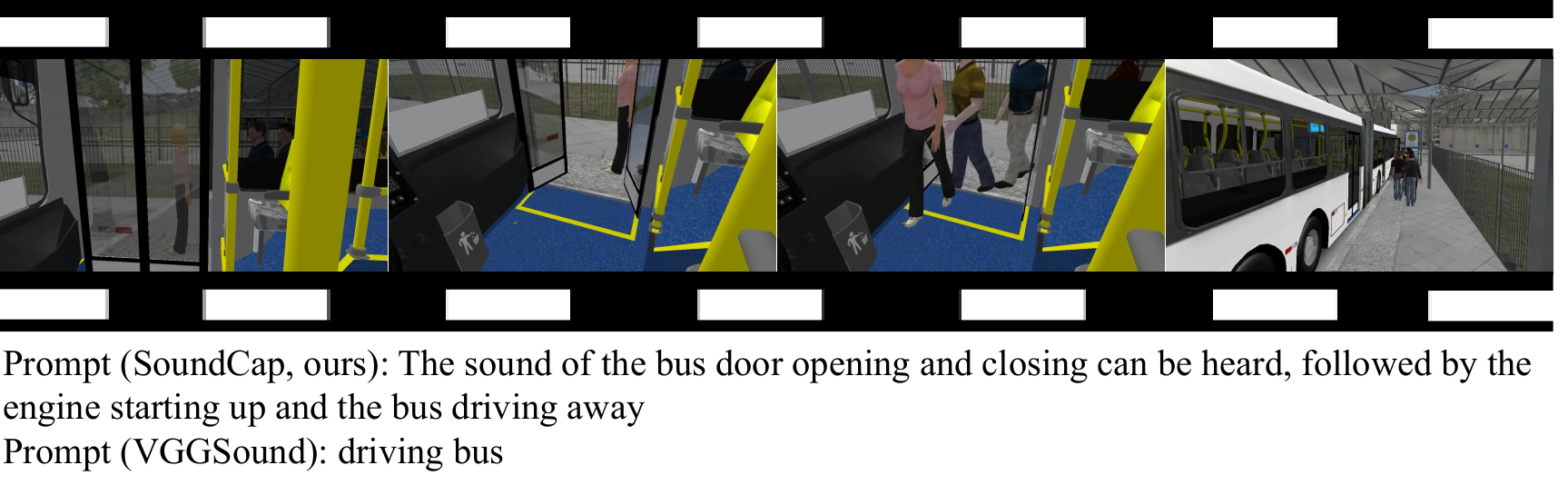}
        \caption{Synthesized caption for video id \texttt{neLRPsl8-XA\_000140}.}
        \label{fig:sub4}
    \end{subfigure}

    \begin{subfigure}[b]{0.49\textwidth}
        \includegraphics[width=\linewidth]{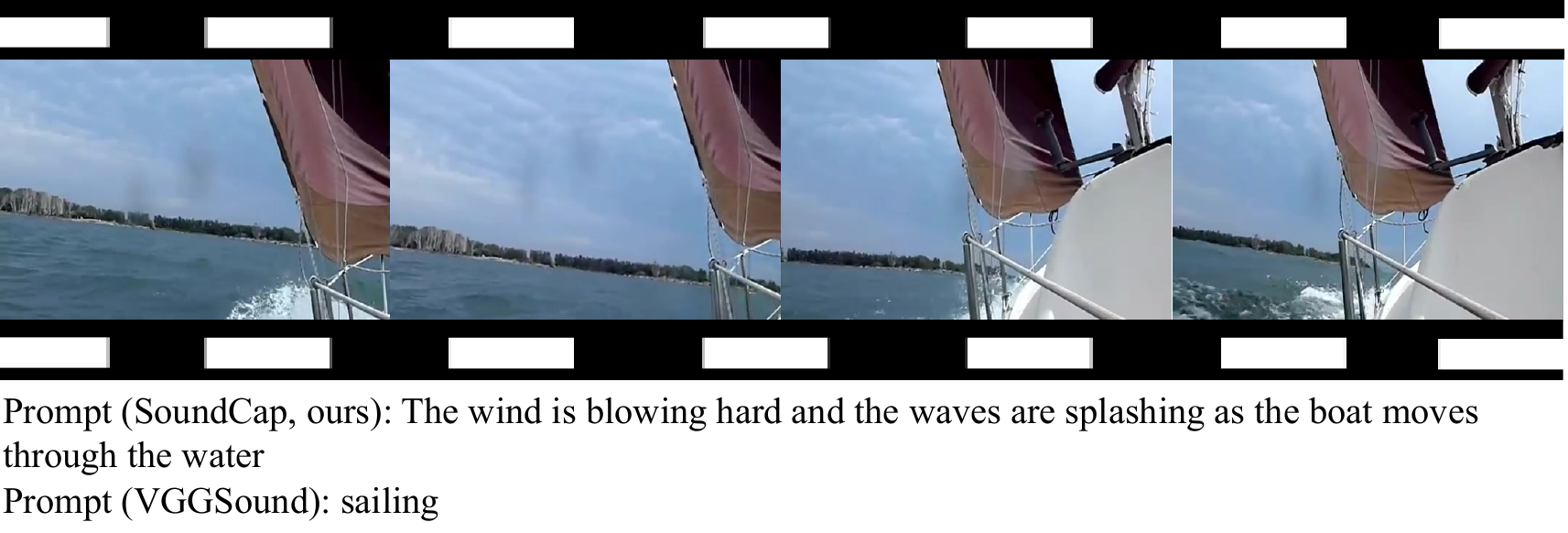}
        \caption{Synthesized caption for video id \texttt{495u-1lHHSc\_000002}.}
        \label{fig:sub5}
    \end{subfigure}
    \hfill
    \begin{subfigure}[b]{0.49\textwidth}
        \includegraphics[width=\linewidth]{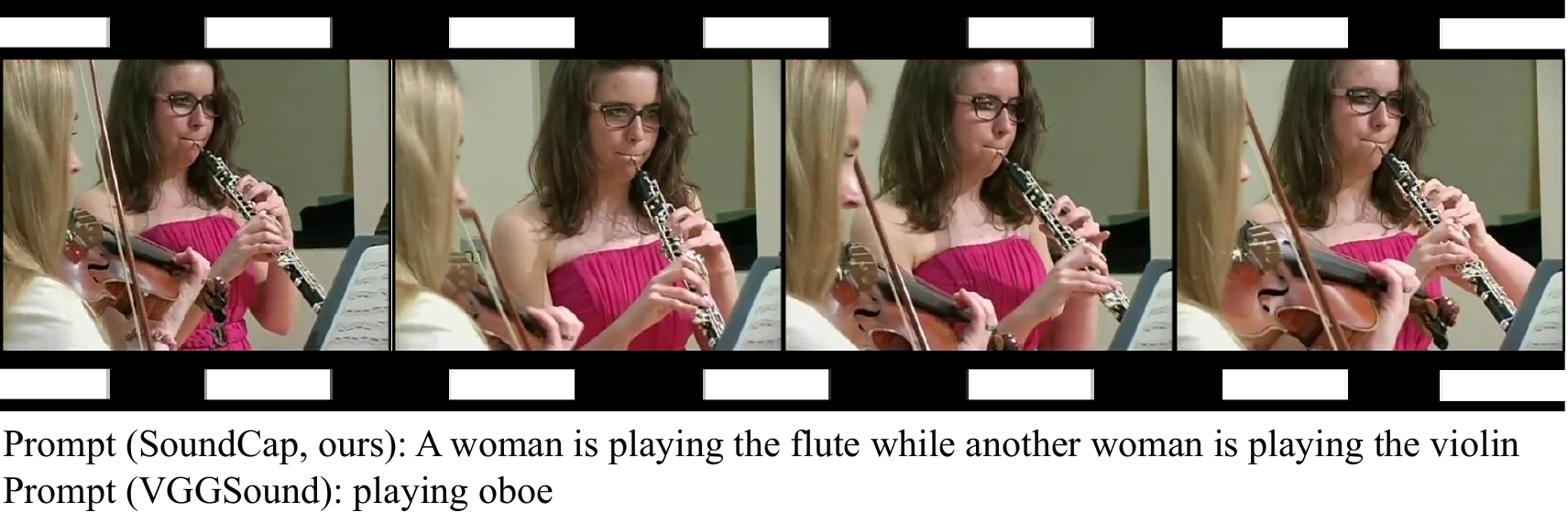}
        \caption{Synthesized caption for video id \texttt{KSx1Y32\_Wd0\_000062}.}
        \label{fig:sub6}
    \end{subfigure}

    \caption{Examples of sound-aware caption generated by SoundCap.}
    \label{fig:soundcap-examples}
\end{figure}

\Cref{fig:soundcap-examples} shows several examples of sound-oriented captions generated by the SoundCap pipeline.

\section{Some Visualizations}\label{abla:qualititive_examples}

\begin{figure}
    \centering
    \includegraphics[width=0.8\linewidth]{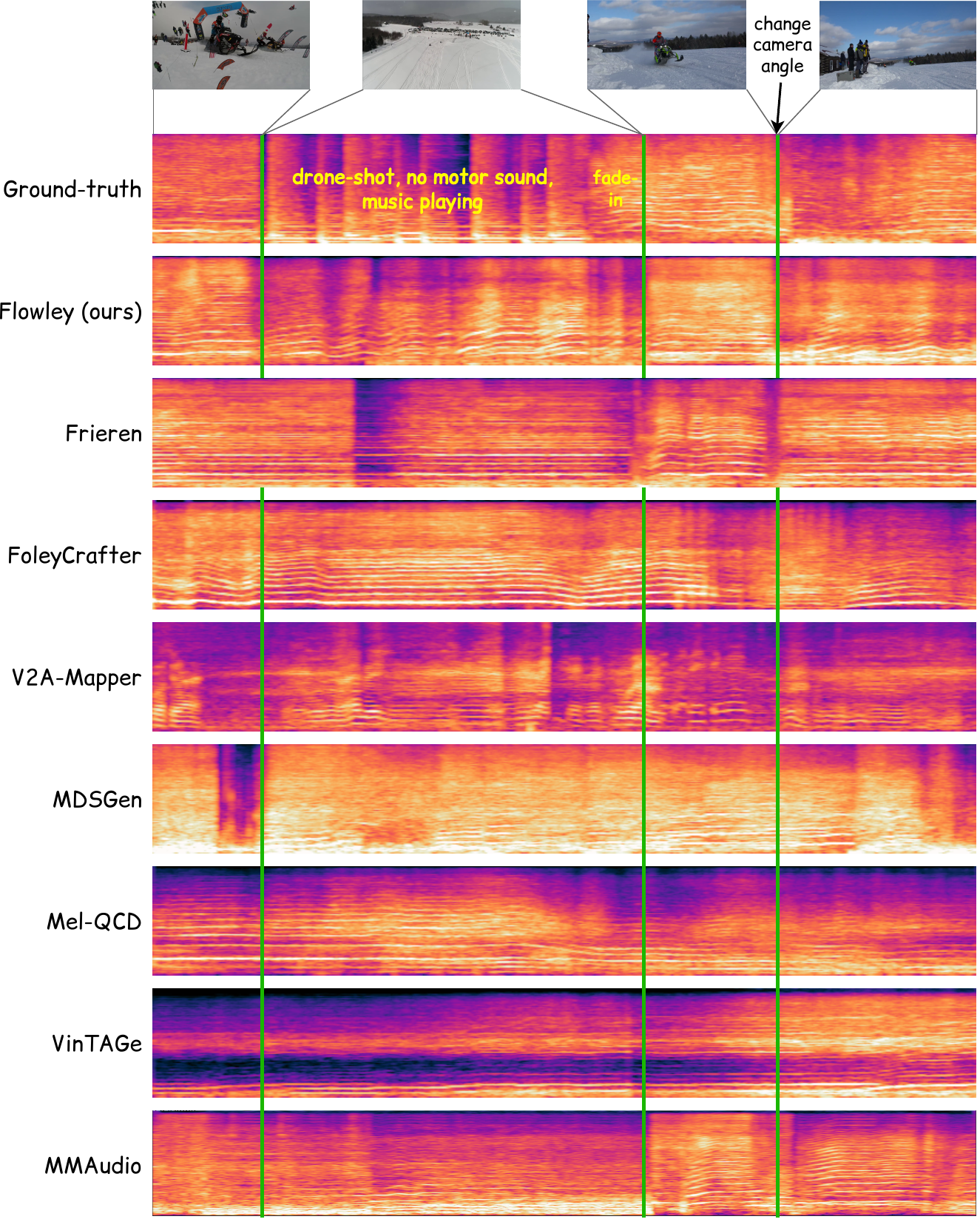}
    \caption{Qualitative comparison on VGGSound. Flowley is the only method that accurately identifies the drone-shot moment, correctly suppressing motor noise. It also captures the transition in camera angle with high precision.}
    \label{fig:spec1}
\end{figure}

\begin{figure}
    \centering
    \includegraphics[width=0.8\linewidth]{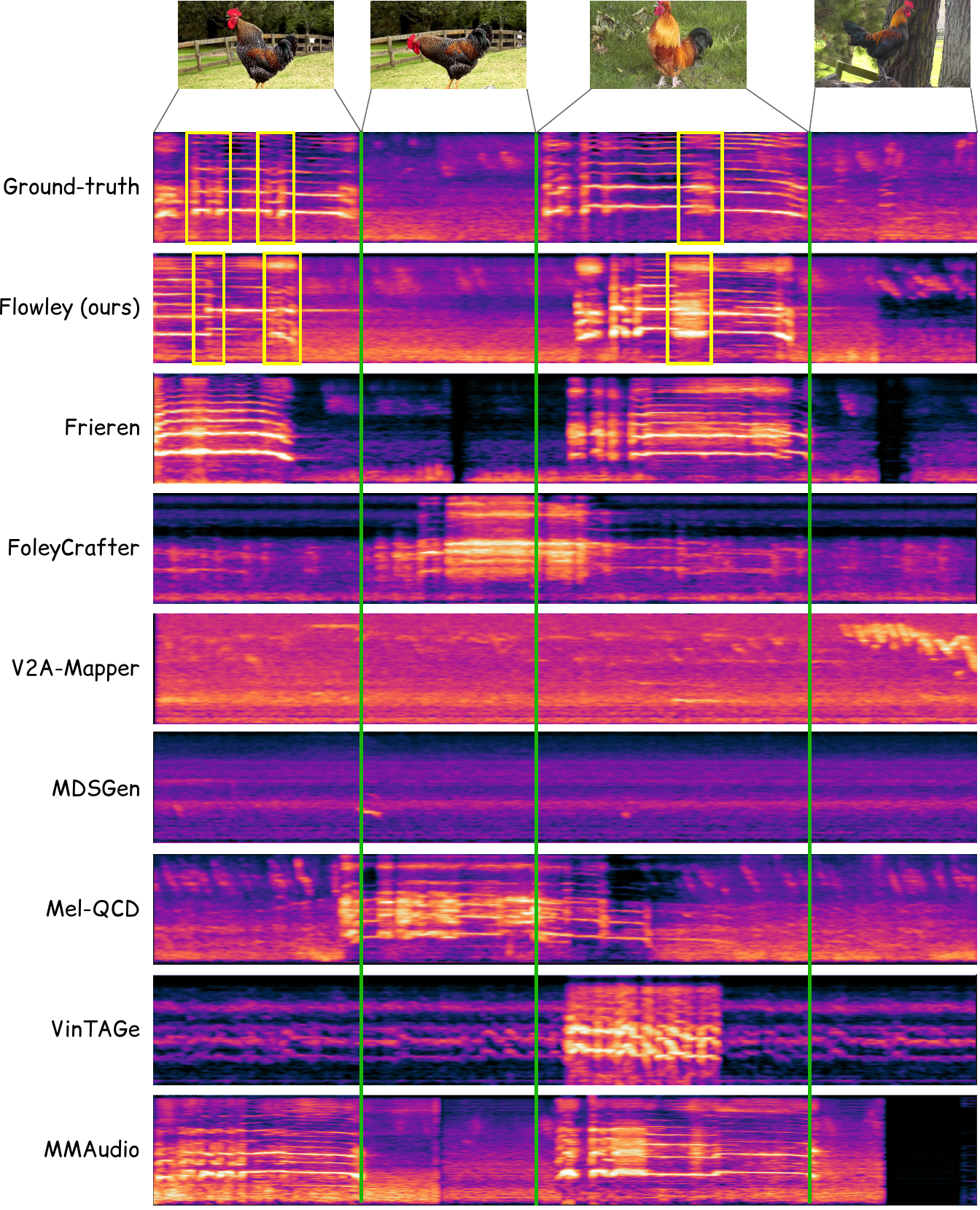}
    \caption{Qualitative comparison on VGGSound. Flowley, Frieren, and MMAudio are the only three methods capable of generating sounds at both distinct crowing moments. However, while the other two fail to accurately capture the timing of the louder, more prolonged crow, Flowley succeeds in doing so, demonstrating notably precise alignment.}
    \label{fig:spec2}
\end{figure}

\begin{figure}
    \centering
    \includegraphics[width=0.8\linewidth]{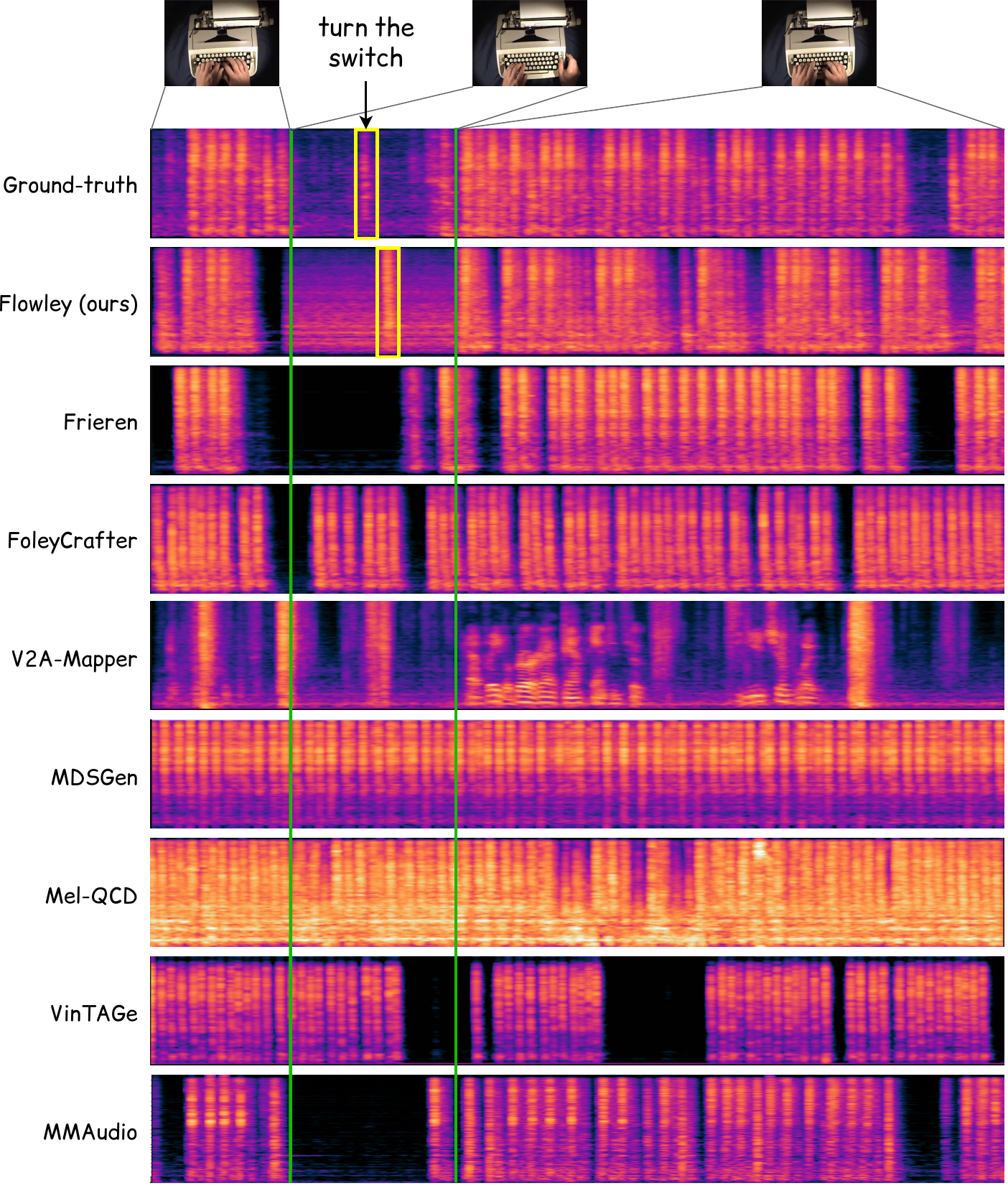}
    \caption{Qualitative comparison on VGGSound. In the fast-motion scenario, although Flowley does not produce audio that is perfectly aligned with the ground-truth, the generated output remains perceptually plausible. Notably, Flowley is the only method that precisely captures the instant when the person flips the switch, producing a subtle ``click'' sound.}
    \label{fig:spec3}
\end{figure}

\begin{figure}
    \centering
    \includegraphics[width=0.8\linewidth]{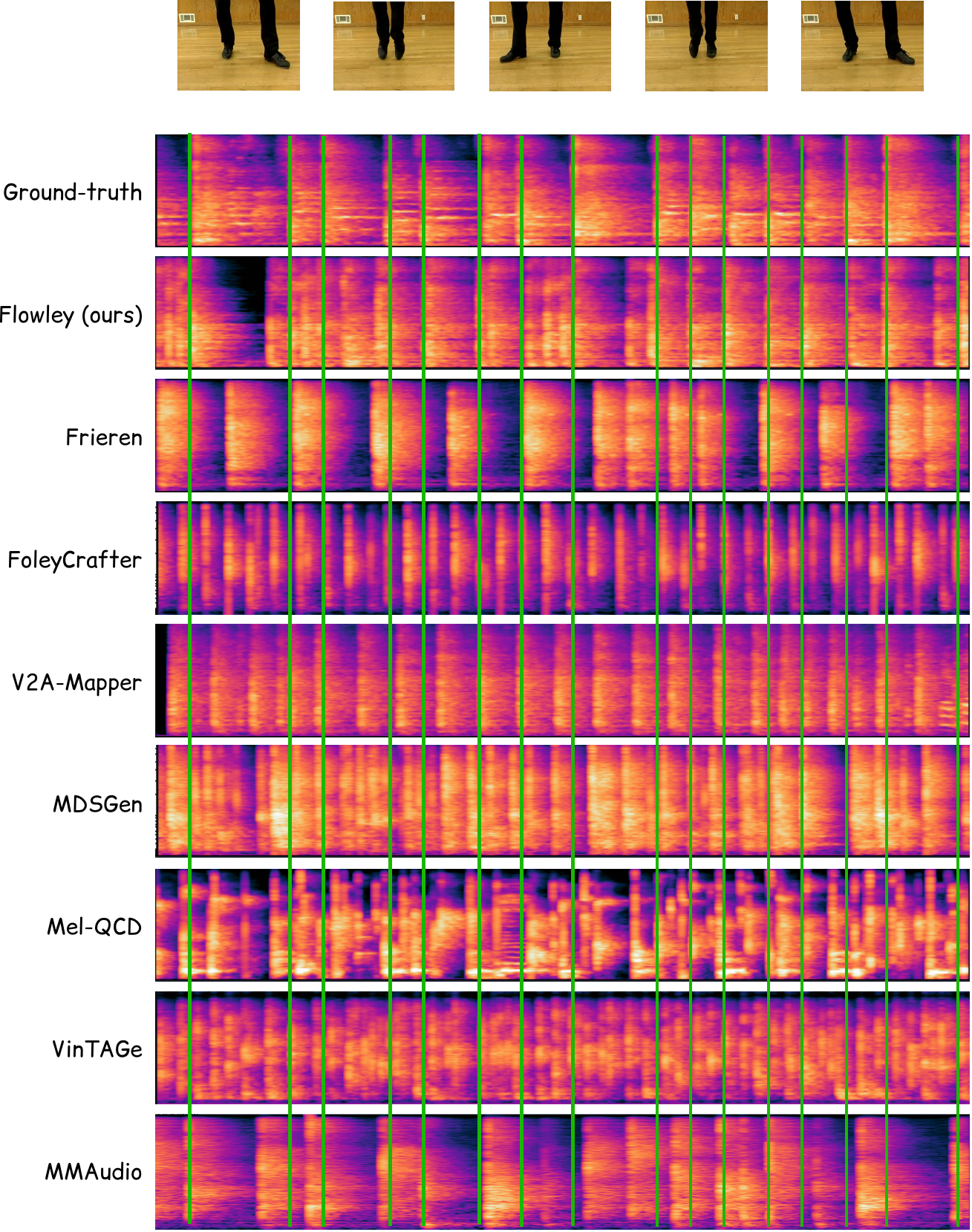}
    \caption{Qualitative comparison on VGGSound. In scenarios with repetitive events (a dancer repeatedly tapping his/her feet on the floor), Flowley still achieves the highest accuracy among all methods.}
    \label{fig:spec4}
\end{figure}

We present several qualitative examples in \Cref{fig:spec1,fig:spec2,fig:spec3,fig:spec4}.

\newpage

\bibliographystyle{splncs04}
\bibliography{main}